\documentclass[11pt]{article}
\usepackage{stmaryrd,amssymb,amsmath,bbm,latexsym,accents,amscd}
\usepackage[all]{xy}
\newtheorem{example}[equation]{Example}
\newtheorem{theorem}[equation]{Theorem}
\newtheorem{corollary}[equation]{Corollary}
\newtheorem{lemma}[equation]{Lemma}
\newtheorem{proposition}[equation]{Proposition}

\numberwithin{equation}{section}

\newcommand{\bigdot}{\accentset{\mbox{\tiny \textbullet}}}
\newcommand{\bigddot}{\accentset{\mbox{\tiny \textbullet\textbullet}}}

\newcommand{\sech}{\hbox{sech}}

\newcommand{\ad}{\hbox{ad}}

\newcommand{\ot}{\otimes}

\newcommand{\fg}{{\mathfrak g}}
\newcommand{\fh}{{\mathfrak h}}
\newcommand{\fm}{{\mathfrak m}}

\newcommand{\gl}{{\mathfrak gl}}
\newcommand{\tr}{\hbox{tr}}

\newcommand{\ga}{\alpha}
\newcommand{\gb}{\beta}
\newcommand{\gs}{\sigma}

\newcommand{\proof}{{\bf Proof}}
\newcommand{\qed}{\hfill $\Box$}

\newcommand{\gd}{\delta}

\newcommand{\ev}{\mathrm{ev}}

\title{Toda systems for Takiff algebras}

\author{Michael Lau\thanks{Funding from the Natural Sciences and
Engineering Research Council of Canada is gratefully acknowledged.} \vspace{0.3cm}
\\{\footnotesize D\'epartement de math\'ematiques et de statistique}\\ {\footnotesize Universit\'e Laval, Qu\'ebec, QC, Canada G1V0A6\  }\\ {\footnotesize Email: Michael.Lau@mat.ulaval.ca\quad\quad\quad\quad\quad\ \ \  }\vspace{0.1cm}}

\begin{document}

\date{}
\maketitle

\begin{small}
\noindent {\bf Abstract:} We study completely integrable systems attached to Takiff algebras $\fg_N$, extending open Toda systems of split simple Lie algebras $\fg$.  With respect to Darboux coordinates on coadjoint orbits $\mathcal{O}$, the potentials of the hamiltonians are products of polynomial and exponential functions.  General solutions for equations of motion for $\fg_N$ are obtained using differential operators called jet transformations.  These results are applied to a $3$-body problem based on $\mathfrak{sl}(2)$, and to an extension of soliton solutions for $A_\infty$ to associated Takiff algebras.  The new classical integrable systems are then lifted to families of commuting operators in an enveloping algebra, solving a Vinberg problem and quantizing the Poisson algebra of functions on $\mathcal{O}$.


\bigskip


\bigskip


\end{small}

\section{Introduction}
Until early computer models suggested otherwise \cite{FPU}, it was widely believed that introducing nonlinearity into particle interactions would result in equipartition of energy among vibrational modes in a lattice.  The hunt for exactly solvable nonlinear lattice models to test this hypothesis led to the discovery of (type $A$) Toda systems \cite{toda67}.  Absence of thermalization was subsequently understood as a consequence of soliton solutions, related to those of the Boussinesq and Korteweg-de Vries equations through approximations of continuum limits of the Toda equations of motion \cite{toda-wadati73}.  Conserved quantitites and Liouville integrability were obtained by Flaschka and H\'enon \cite{flaschka, henon}, and later extended to open Toda systems of all simple types by Kostant \cite{kostant79}.

In this paper, we introduce and study a new family of completely integrable systems attached to Takiff algebras $\fg_N=\fg\ot_K K[u]/\langle u^{N+1}\rangle$.  Takiff algebras are finite dimensional quotients of current Lie algebras $\fg\ot_K K[u]$ for split simple Lie algebras $\fg$ over the real or complex field $K$.  For example, $\fg_N$ is the simple Lie algebra $\fg$ when $N=0$, and when $N=1$, $\fg_N$ is an extension of $\fg$ by its adjoint representation.  These Lie algebras are neither semisimple nor reductive when $N>0$, but nonetheless have nondegenerate symmetric invariant bilinear forms.  This easy but crucial observation identifies them with their duals, and lets us define integrable systems on coadjoint orbits using a Lax approach.

Takiff algebras have appeared many times in recent mathematical physics literature, for example, in \cite{LT}, where they are used to construct a finite dimensional Lie group governing the Bloch-Iserles equation, in \cite{BR}, where their non-semisimplicity is exploited to find indecomposable representations that are not irreducible, in the context of logarithmic conformal field theory, in \cite{Casati-Ortenzi06}, where they are used to study coupled Hirota bilinear equations, and in \cite{MY}, where they appear as Lie algebras of jet groups, later used to study associated varieties of affine $W$-algebras \cite{arakawa}.

We deform the Lie bracket of $\fg_N$ using a classical $r$-matrix satisfying the modified Yang-Baxter equation, and consider distinguished coadjoint orbits $\mathcal{O}$ of the corresponding Lie groups.  The orbits $\mathcal{O}$ are symplectic manifolds parametrized by Lax matrices.  In contrast with the split simple Lie algebra case, these matrices no longer have enough eigenvalues to generate an integrable system.  To resolve this difficulty, we enlarge the supply of functions by replacing ordinary traces of matrices with traces along superdiagonals in a faithful representation.  This leads to a classification of invariant bilinear forms on $\fg_N$, and lets us construct a maximal independent family $\mathbb{I}_N(\fg)$ of Poisson-commuting functions on $\mathcal{O}$, generalizing the Toda systems for finite dimensional split simple Lie algebras (Theorem \ref{system-is-integrable-thm}).

Equations of motion (\ref{q_i-dot}) and (\ref{p_i-dot}) are then explicitly derived in terms of position and momentum coordinates on coadjoint orbits, using exponential generating functions defined in Proposition \ref{darboux-defn-lemma}.  We concentrate on evolution equations governed by hamiltonians with the usual quadratic expression for kinetic energy, as the most physically relevant members of new integrable hierarchies.  The corresponding potentials are products of polynomial and exponential functions and have not, to the best of our knowledge, previously appeared in the literature.  Standard techniques based on factorisation in Lie groups are then used to exactly solve a minimal rank example.

In Section \ref{jet-transformations-subsection}, we introduce differential operators $D_n$ called jet transformations.  Applied to any solution of the Toda equations of motion for a split simple Lie algebra $\fg$, the jet transformations give solutions to the equations attached to the Takiff algebras $\fg_N$.  Since general solutions of classical Toda equations are known for all split simple Lie algebras $\fg$ by \cite{kostant79}, jet transformations provide general solutions for all Takiff algebras $\fg_N$.  We illustrate with the general solution of a $3$-body problem based on $\mathfrak{sl}(2)$ and an extension of soliton solutions for the $A_\infty$-lattice to Takiff algebras.  

The final section considers the Vinberg problem of identifying commutative subalgebras of enveloping algebras, corresponding to the new integrable systems $\mathbb{I}_N(\fg)$ studied in this paper.  This is done with a Harish-Chandra projection and $r$-deformation of an analogous construction \cite{molev} of generators for centres of enveloping algebras of Takiff algebras.  Conserved quantities in $\mathbb{I}_N(\fg)$ then appear as principal symbols of images of the generators.  Theorem \ref{quantum-involution-thm} can thus be seen as a quantization of $\mathbb{I}_N(\fg)$ in an enveloping algebra whose associated graded algebra is the ring of functions on $\mathcal{O}$.

\bigskip

\setcounter{section}{1}

\section{Takiff algebras}

Let $\fg$ be a nonzero finite-dimensional Lie algebra over a field $K$ of characteristic zero.  Without loss of generality, we fix a faithful representation $\rho:\ \fg\rightarrow \mathfrak{gl}(M)$ of minimal dimension $M$, and identify $\fg$ with its image in the general linear algebra $\mathfrak{gl}(M)$.  For each nonnegative integer $N$, let
$$\fg_N=\fg\ot_K\left(K[u]/\langle u^{N+1}\rangle\right)$$
be the {\em Takiff algebra}, or {\em truncated current algebra}, of degree $N$, where $\langle u^{N+1}\rangle$ is the principal ideal generated by $u^{N+1}$ in the polynomial ring $K[u]$.  We write $x(i)$ for the image $x\ot(u^i+\langle u^{N+1}\rangle)$ of $x\ot u^i$ in $\fg_N$.  The Lie bracket on $\fg_N$ is given by linear extension of the bracket on $\fg$:
$$[x(i),y(j)]=[x,y](i+j),$$
keeping in mind that $x(i)=0$ for $i>N$.  Note that $\fg_N$ generalizes $\fg=\fg_0$ and is a quotient of the current algebra $\fg\ot_K K[u]$.

The Lie algebra $\fg_N$ embeds in $\mathfrak{gl}((N+1)M)$ as a collection of matrices of diagonally repeating $M\times M$ blocks:

\begin{eqnarray*}x_0+x_1(1)+\cdots+x_N(N)\hookrightarrow\left(\begin{array}{llll}
    x_0 & x_1 & \cdots & x_N\\
    0 & x_0 & \cdots & x_{N-1}\\
    \vdots & \vdots & \ddots & \vdots\\
    0 & 0 & \cdots & x_0
    \end{array}\right),
\end{eqnarray*}
for all $x_0,\ldots , x_N\in\fg\subseteq \mathfrak{gl}(M)$.  By convention, the superdiagonals of such matrices in $\mathfrak{gl}((N+1)M)$ will be numbered from $0$ to $(N+1)M-1$, with the principal diagonal as the $0$th superdiagonal.  For $A\in\gl((N+1)M)$, the trace along the $\ell M$th superdiagonal will be denoted by $\tr_\ell(A)$, so that
$$\tr_\ell(x(i))=(N+1-\ell)\gd_{i\ell}\tr(x),$$
for all $x\in\fg$ and $0\leq i\leq N$.  Note that $\tr_0$ is the ordinary trace of a matrix in $\gl((N+1)M)$.

When $N>0$, the Lie algebra $\fg_N$ is never reductive, but nonetheless has a symmetric invariant bilinear form $(-|-)_N$ which is nondegenerate whenever $\fg$ is semisimple:
$$(x(i)|y(j))_N=\gd_{i+j,N}\tr(xy)=\tr_N(x(i)y(j)),\ \hbox{for all\ }x,y\in\fg\ \hbox{and}\ i,j=0,\ldots, N,$$
where $xy$ and $x(i)y(j)$ are the associative products of $M\times M$ and $(N+1)M\times (N+1)M$ matrices, respectively.  This observation was the starting point for our study of Takiff algebras, since it lets us identify them with their duals and explore Toda systems in a non-reductive setting.  We write
$\nu:\ \fg_N\rightarrow \fg_N^*$ for the induced isomorphism, where $\nu(x)(y)=(x|y)_N$ for all $x,y\in \fg_N$.  More generally,

\begin{lemma}{\bf \cite[Theorem 2.2]{Casati-Ortenzi06}}
  If $(-,-)_\fg$ is a nondegenerate symmetric invariant bilinear form on $\fg$ and $c=(c_0, \ldots , c_N)\in K^{N+1}$, then
  $$(x(i),y(j))_c=c_{i+j}(x,y)_\fg, \hbox{\ for all\ }x,y\in\fg\hbox{\ and\ }0\leq i,j\leq N$$
  defines a symmetric invariant bilinear form on $\fg_N$.  The form $(-,-)_c$ is nondegenerate if and only if $c_N\neq 0$.\qed
  \end{lemma}
Such forms $(-,-)_c$ are often the only possibility:
\begin{lemma} If $\fg$ is simple and $K$ is algebraically closed, then for any symmetric invariant bilinear form $(-,-):\ \fg_N\times\fg_N\rightarrow K$, there exist $c_0,\ldots, c_N\in K$ so that $(x(i),y(j))=c_{i+j}\tr(xy)$ for all $x,y\in\fg$ and $0\leq i,j\leq N$.  For $\ell=0,\ldots, N$, the forms $(\ |\ )_\ell$ defined by
  $$(x(i)|y(j))_\ell=\gd_{i+j,\ell}\tr(xy)=\frac{1}{N+1-\ell}\tr_\ell(x(i)y(j))$$
  are thus a basis for the vector space of all symmetric invariant bilinear forms on $\fg_N$.
  \end{lemma}
\proof\ \ Let $0\leq i,j\leq N$ and define a bilinear map $\lambda_i:\ \fg\times\fg\rightarrow K$ by
$$\lambda_i(x,y)=(x(0),y(i)),$$
for all $x,y\in\fg$.  For all $x,y,z\in\fg$,
\begin{align*}
  \lambda_i(x,[y,z])&=(x(0),[y(0),z(i)])\\
  &=(y(0),[z(i),x(0)])\\
  &=(y(0),[z(0),x(i)])\\
  &=([y(0),z(0)],x(i))\\
  &=\lambda_i([y,z],x),
  \end{align*}
and since $[\fg,\fg]=\fg$, this shows that $\lambda_i$ is symmetric.  A similar easy calculation shows that $\lambda_i$ is also invariant.  Finally, $\fg$ is simple and $K$ is algebraically closed, so every symmetric invariant bilinear form on $\fg$ is a multiple of the trace form (itself a normalization of Killing form), and there exist $c_0,\ldots,c_N\in K$ such that $\lambda_i(x,y)=c_i\tr(xy)$ for all $x,y\in\fg$.

This completes the proof since $[\fg,\fg]=\fg$ and
\begin{align*}
  ([x,y](i),z(j))&=(x(0),[y(i),z(j)])\\
  &=\lambda_{i+j}(x,[y,z])\\
&=\lambda_{i+j}([x,y],z)\\
  &=c_{i+j}\tr([x,y],z).
  \end{align*}
\qed

\bigskip

That the forms $(\ |\ )_\ell$ are clearly symmetric will later be used in the proof of integrability in Section 3.3.


\section{Takiff Toda systems and integrability}

In this section, $\fg$ will be a finite dimensional split simple Lie algebra with Cartan subalgebra $\fh$, root system $\Phi$, and root spaces $\fg_\ga=\{x\in \fg\ :\ [h,x]=\ga(h)x\hbox{\ for all\ }h\in\fh\}$, over the field $K$ of real or complex numbers.  We will give our proofs in the real context, as this is the usual environment for applications; analogous proofs hold in the complex setting.  Fix a base $\Delta=\{\ga_1,\ldots,\ga_s\}$ of simple roots, and denote the corresponding sets of positive and negative roots by $\Phi_+$ and $\Phi_-$, respectively.  As in the previous section, we view $\fg$ as a matrix Lie algebra by fixing a faithful representation of minimal dimension $M$.  Let $\{h_1,\ldots,h_s\}$ be an orthonormal basis for $\fh$, with respect to the trace form on $\fg$.  Choose nonzero root vectors $e_\ga\in\fg_\ga$ and $f_\ga\in\fg_{-\ga}$ for each positive root $\ga$, normalized so that $\tr(e_\ga f_\ga)=1$.  Then $[e_\ga,f_\ga]=h_\ga$, where $h_\ga=\sum_i \ga(h_i)h_i.$  All Lie algebras, Lie groups, bilinear forms, tensor products, and spans will be taken over $K$.  As usual, the integers will be denoted by $\mathbb{Z}$, and the nonnegative integers by $\mathbb{Z}_+$.

\subsection{Lie bialgebra structure}

For each $X\in\fg_N\subseteq \gl((N+1)M)$ and $Y=\sum_i a_i\ot b_i\in\fg_N\ot \fg_N$, we use the physics notation $X_1=X\ot 1,\ X_2=1\ot X\in\gl((N+1)M)\ot\gl((N+1)M)$, and $Y_{12}=Y,\ Y_{21}=\sum b_i\ot a_i\in\fg_N\ot \fg_N$.  When the meaning is clear from the context, we will abuse notation and also write
$$Y_{12}=Y\ot 1,\ Y_{13}=\sum_i a_i\ot 1\ot b_i,\hbox{\ and\ }Y_{23}=1\ot Y,$$
as elements of $\gl((N+1)M)^{\ot 3}$.  The block partial trace operators $\hbox{Tr}_{\ell,j}$ are defined by taking the block trace $\tr_\ell$ on the $j$th tensor component:
\begin{align*}
  \hbox{Tr}_{\ell,1}(Y)&=\sum_i\tr_\ell(a_i)b_i\\
  \hbox{Tr}_{\ell,2}(Y)&=\sum_i\tr_\ell(b_i)a_i.
  \end{align*}

In a generalisation of the usual Toda $r$-matrix, we define
\begin{equation}\label{r-matrix}
r_{12}=\frac12\sum_{\ga\in\Phi_+}\sum_{i=0}^N(e_\ga(i)\ot f_\ga(N-i)-f_\ga(N-i)\ot e_\ga(i))\in\fg_N\ot\fg_N.
\end{equation}
The $2$-tensor $r_{12}$ defines a Rota-Baxter operator $R:\ \fg_N\rightarrow\fg_N$ by $R(X)=\hbox{Tr}_{N,2}(r_{12}X_2).$  Calculating directly,
$$R(h_i(j))=0,\ R(e_\ga(j))=\frac12 e_\ga(j),\ R(f_\ga(j))=-\frac12 f_\ga(j),$$
for all $\ga\in\Phi_+,\ 1\leq i\leq s,$ and $0\leq j\leq N$.  The resulting $R$-deformed bracket
$$[X,Y]_R=[RX,Y]+[X,RY]$$
satisfies the modified Yang-Baxter equation
$$[RX,RY]-R([X,Y]_R)=-\frac14[X,Y]$$
for all $X,Y\in\fg_N$.  It then follows that $\fg_N$ is a Lie bialgebra, equipped with brackets $[-,-]$ and $[-,-]_R$.  See \cite{Babelon-Bernard-Talon03} or \cite{Chari-Pressley94} for more details on Lie bialgebras and the modified Yang-Baxter equation.  We will write $\fg_N$ for the Lie algebra $(\fg_N,[-,-])$ and $\fg_N^R$ for $(\fg_N,[-,-]_R)$.  Explicitly,
\begin{align*}
  [h_i(k),h_j(\ell)]_R&=0\\
  [h_i(k),e_\ga(\ell)]_R&=\frac12\ga(h_i)e_\ga(k+\ell)\\
[h_i(k),f_\ga(\ell)]_R&=\frac12\ga(h_i)f_\ga(k+\ell)\\
[e_\ga(k),e_\gb(\ell)]_R&=[e_\ga,e_\gb](k+\ell)\\
[e_\ga(k),f_\gb(\ell)]_R&=0\\
[f_\ga(k),f_\gb(\ell)]_R&=-[f_\ga,f_\gb](k+\ell),
  \end{align*}
for all $1\leq i,j\leq s,\ 0\leq k,\ell\leq N,$ and $\ga,\gb\in\Phi_+$.  The Lie algebra $\fg_N^R$ is thus solvable and has a weight space decomposition with respect to the adjoint action of $\fh^R=\hbox{Span}\{h_i(0)\ :\ 1\leq i\leq s\}$:
$$(\fg_N^R)_\ga=\hbox{Span}\{e_\ga(i),f_\ga(i)\ :\ 0\leq i\leq N\}$$
  for all $\ga\in\Phi_+$.
  \subsection{Poisson structure}
  The symmetric algebra $S(\fg_N)$ of regular functions on the algebraic dual $\fg_N^*$ of $\fg_N$ has natural Poisson brackets $\{-,-\}$ and $\{-,-\}_R$ induced from the bialgebra structure of $\fg_N$, defined by the Leibniz rule and relations
 \begin{eqnarray}
    \{x,y\}&=&[x,y]\nonumber\\
    \{x,y\}_R&=&[x,y]_R\hbox{\ for all\ }x,y\in\fg_N.\label{R-bracket-defn}
    \end{eqnarray}

  Let $G_N$ and $G_N^R$ be connected Lie groups with Lie algebras $\fg_N$ and $\fg_N^R$, respectively.  To obtain a symplectic structure and extend known Toda systems, we will consider functions on certain $G_N^R$-coadjoint orbits
  $$\mathcal{O}_\lambda=(\hbox{Ad}_R^*G_N^R)\lambda\subset \fg_N^*.$$
  Such orbits will be identified with their preimages
  $$\mathcal{O}_{x_\lambda}=(\hbox{Ad}_R^*G_N^R)x_\lambda=\nu^{-1}(\mathcal{O}_{\lambda})\subset\fg_N$$
  under the linear isomorphism
  \begin{eqnarray*}
    \nu:\ \fg_N&\longrightarrow&\fg_N^*\\
    x&\longmapsto&(x|-)_N,
  \end{eqnarray*}
 where $x_\lambda=\nu^{-1}(\lambda)$.  We will refer to both $(\fg_N^*,\ \hbox{Ad}_R^*)$ and $(\fg_N,\hbox{Ad}_R^*)$ as the {\em coadjoint representation} of $G_N^R$, where the coadjoint action on $\fg_N$ is given by $(\hbox{Ad}_R^*g)x=\nu^{-1}((\hbox{Ad}_R^*g)\nu(x))$, for all $g\in G_N^R$ and $x\in\fg_N$.  The definition of the coadjoint action $\hbox{ad}_R^*$ of $\fg_N^R$ on the vector space $\fg_N$ is made analogously. 

  In contrast with the usual Poisson set-up, the form $(-|-)_N$ is not $G_N^R$-invariant, so the coadjoint action of $\fg_N^R$ is {\em not} equivalent to its adjoint action under the linear isomorphism $\nu$.  Rather, for $x,y,z\in\fg_N$ and $\mu=\nu(y)\in\fg_N^*$, we have
  \begin{align*}
    (\ad_R^*(x)\mu)z&=(y|[Rz,x])_N+(y|[z,Rx])_N\\
    &=([x,y]|Rz)_N+([Rx,y]|z)_N.
    \end{align*}
  As the $2$-tensor $r_{12}$ is skew-symmetric, the operator $R$ is skew-adjoint, and
  $$([x,y]|Rz)_N=(-R[x,y]|z)_N.$$
  The induced coadjoint action on $\fg_N=\nu^{-1}(\fg_N^*)$ is thus
  $$(\ad_R^*x)y=-R[x,y]+[Rx,y],$$
  for all $x,y\in\fg_N$.  Explicitly,
  \begin{align}
    \ad_R^*(\fg_N)h_i(\ell)&=0 \nonumber\\
    \ad_R^*(h_i(k))e_\gb(\ell)&=-\frac12\beta(h_i)e_\gb(k+\ell)\nonumber\\
    \ad_R^*(h_i(k))f_\gb(\ell)&=-\frac12\beta(h_i)f_\gb(k+\ell)\nonumber\\
    \ad_R^*(e_\ga(k))f_\gb(\ell)&=\left\{\begin{array}{ll}
    \frac12h_\ga(k+\ell) & \hbox{if}\ \ga=\gb\\
     \lbrack e_\ga,f_\gb\rbrack(k+\ell) & \hbox{if}\ \gb-\ga\in\Phi_+\\
     0 & \hbox{otherwise}\nonumber\end{array}\right.\\
\ad_R^*(f_\ga(k))e_\gb(\ell)&=\left\{\begin{array}{ll}
    \frac12h_\ga(k+\ell) & \hbox{if}\ \ga=\gb\\
     \lbrack e_\gb,f_\ga\rbrack(k+\ell) & \hbox{if}\ \gb-\ga\in\Phi_+\\
     0 & \hbox{otherwise}\nonumber\end{array}\right.\\
\ad_R^*(e_\ga(k))e_\gb(\ell)&=\ad_R^*(f_\ga(k))f_\gb(\ell)=0\label{coadjoint-formulas},
    \end{align}
  for all $\ga,\gb\in\Phi_+$, $1\leq i\leq s$, and $0\leq k,\ell\leq N.$

  \medskip
  
  \begin{proposition}\label{coadjoint-orbit-description-proposition} Let $V=\hbox{Span}\{h_i(k),e_\ga(k)+f_\ga(k)\,:\,\ga\in\Delta,\ 1\leq i\leq s,\ 0\leq k\leq N\}$ and $x'=\sum_{\ga\in\Delta}(e_\ga(0)+f_\ga(0))$.
    \begin{enumerate}
    \item[{\rm (i)}] The space $V$ is a $G_N^R$-submodule of the coadjoint representation.
    \item[{\rm (ii)}] The coadjoint orbit $\mathcal{O}=\hbox{Ad}_R^*(G_N^R)x'$ is an open subset of the linear space $V$.  In particular, the Poisson algebra of polynomial functions on $\mathcal{O}$ can thus be identified with the symmetric algebra $(S(V),\{-,-\}_R)$.
      \end{enumerate}    
    \end{proposition}

  \noindent
  \proof \ \ (i)\ Indeed, when $\ga\in\Phi_+$ and $\gb\in\Delta$ in (\ref{coadjoint-formulas}), the condition $\gb-\ga\in\Phi_+$ is never satisfied, so $V$ is stable under $\ad_R^*(\fg_N)$.  It is thus a submodule of the coadjoint representation for both $\fg_N^R$ and $G_N^R$.

  \medskip
  
  \noindent
  (ii) By Part\,(i), $\mathcal{O}\subseteq V$.  The fact that $\mathcal{O}$ is open follows from the inverse function theorem.  In particular, the exponential is a local diffeomorphism between an open neighbourhood of $0\in\fg_N^R$ and a neighbourhood $U$ of the identity element $1\in G_N^R$.  The coadjoint action of each $g\in U$ is thus the exponential of the action of the corresponding element of $\fg_N^R$.  The differential of
  \begin{eqnarray*}
    \phi: \ G_N^R &\longrightarrow&V\\
    g &\longmapsto&\hbox{Ad}_R^*(g)x'
    \end{eqnarray*}
  at $1\in G_N^R$ is then the linear map
  \begin{eqnarray*}
    \rho: \ \fg_N^R &\longrightarrow&V\\
    y &\longmapsto&\hbox{ad}_R^*(y)x',
    \end{eqnarray*}
  whose kernel is the annihilator $\fg_N^{x'}$ of $x'$.

  By Formulas (\ref{coadjoint-formulas}), we see that
  $$\fg_N^{x'}=\hbox{Span}\{e_\ga(k)-f_\ga(k),\ e_\gb(k),\ f_\gb(k)\,:\,\ga\in\Delta,\ \gb\in\Phi\setminus\Delta,\ 0\leq k\leq N\},$$
  a vector space of dimension
  $$(N+1)\big(\hbox{card}(\Delta)+2\hbox{card}(\Phi\setminus\Delta)\big)=(N+1)(\dim\fg-2s).$$
  The image of $\rho$ is therefore of dimension
  $$\dim(\fg_N/\fg_N^{x'})=\dim\fg_N-\dim\fg_N^{x'}=2s(N+1)=\dim V,$$
  and $\rho$ is surjective.

  It now follows from the inverse function theorem that $\phi$ is an open map when restricted to a sufficiently small open neighbourhood $U'$ of $1$.  The subset $\phi(U')=\hbox{Ad}_R^*(U')x'$ is thus open.  The group $G_N^R$ acts by diffeomorphisms on $V$, so $\hbox{Ad}_R^*(g)\phi(U')\subset V$ is open for all $g\in G_N^R$.  Therefore, $\mathcal{O}_{x'}=\bigcup_{g\in G_N^R}\hbox{ad}_R^*(g)\phi(U')\subset V$ is also open.\qed

\bigskip
  
  The Poisson structure $\{-,-\}_R$ on $\fg_N^*$ induces a symplectic structure on $\mathcal{O}$ and thus on $V$, via the Kostant-Kirillov-Souriau $2$-form
  $$\omega_\lambda(\ad_R^*(y)x',\ad_R^*(z)x')=\lambda[y,z],$$
  for $\lambda=\nu(x')$ and any $y,z\in\fg_N^R$.  As usual, this form is well defined, nondegenerate, and closed.

  \subsection{Integrable systems}
  
Recall that a {\em (Liouville) integrable system} is a symplectic manifold $(M,\omega)$ of dimension $2n$, together with $n$ independent (smooth or holomorphic) functions $f_1,\ldots, f_n$ that commute with each other and with a hamiltonian function $H$ on $M$, relative to the Poisson bracket $\{f,g\}=\omega(\xi_f,\xi_g)$, where $\hbox{d}f=\omega(\xi_f,-)$.  {\em Independent} means that $\hbox{d}f_1\wedge\cdots\wedge\hbox{d}f_n\neq 0$ on a dense open subset of $M$.  Note that independence and algebraic independence are equivalent in contexts where $f_1,\ldots, f_n$ are polynomial functions.

Integrability for Toda systems associated to finite dimensional split simple Lie algebras was proved by Kostant \cite{kostant79}, following earlier work by Flaschka and H\'enon \cite{flaschka, henon} in type $A$.  Such integrable systems have a Lax presentation, where the Lax matrix $L$ parametrizes (the dual of) a $G_0^R$-coadjoint orbit.  Entries in $L$ are thus dynamic (time-dependent) coefficient functions on the orbit.  For finite dimensional simple Lie algebras, the eigenvalues of the Lax matrix, or equivalently, the traces of powers of $L$, supply the necessary conserved quantities $f_1,\ldots , f_n$ relative to the Toda hamiltonian $H=\frac12\tr(L^2)$.

In contrast with the split simple Lie algebra setting, Lax matrices for Takiff algebras no longer have enough eigenvalues to generate an integrable system.  For example, when $\fg=\mathfrak{sl}(2)$, the natural Lax matrix parametrizing the $G_N^R$-coadjoint orbit $\mathcal{O}=\mathcal{O}_{x'}$ of Proposition \ref{coadjoint-orbit-description-proposition} is
$$L=\left(\begin{array}{llll}
    x_0 & x_1 & \cdots & x_N\\
    0 & x_0 & \cdots & x_{N-1}\\
    \vdots & \vdots & \ddots & \vdots\\
    0 & 0 & \cdots & x_0
    \end{array}\right),$$
where $x_k$ is the $2\times 2$ matrix
$$\left(\begin{array}{cc}
\frac{1}{\sqrt{2}}y(k)&b(k)\\
b(k)&-\frac{1}{\sqrt{2}}y(k)\end{array}\right)\in\fg,$$
and $y(k),b(k):\mathcal{O}\rightarrow K$ are coefficient functions of basis elements $h_1(k)=\frac{1}{\sqrt{2}}h_\ga(k)$ and $e_\ga(k)+f_\ga(k)$, respectively, in the expressions
$$\sum_{i=0}^Ny(k)h_1(k)+b(k)(e_\ga(k)+f_\ga(k))\in\mathcal{O},$$
where $\Delta=\{\ga\}$.  Then
\begin{equation}\label{star}
\tr(L^n)=\left\{\begin{array}{cl}
\left(\frac12 y(0)^2+b(0)^2\right)^{n/2}&\hbox{if\ }n\hbox{\ is even}\\
0&\hbox{if\ }n\hbox{\ is odd}.
\end{array}\right.
\end{equation}
The differentials of the functions $\tr(L^n)$ are thus linearly dependent everywhere on the $2(N+1)$-dimensional symplectic manifold $\mathcal{O}$.

The solution to the paucity of independent functions in (\ref{star}) is to enlarge the definition of the trace to include the block traces $\tr_\ell$.  With this in mind, for any finite dimensional split simple Lie algebra $\fg$, let
$$L=\left(\begin{array}{llll}
    x_0 & x_1 & \cdots & x_N\\
    0 & x_0 & \cdots & x_{N-1}\\
    \vdots & \vdots & \ddots & \vdots\\
    0 & 0 & \cdots & x_0
    \end{array}\right),$$
where $x_k$ is the submatrix
$$x_k=\sum_{i=1}^sy_i(k)h_i+\sum_{\ga\in\Delta}b_\ga(k)(e_\ga+f_\ga)\in\fg\subset\mathfrak{gl}(M),$$
and $y_i(k), b_\ga(k)$ are the corresponding coefficient functions on $\mathcal{O}=\mathcal{O}_{x'}$.  Using the definition of the $R$-bracket (\ref{R-bracket-defn}) and calculating directly on $\mathcal{O}$, we see that
\begin{eqnarray}
\{y_i(k),y_j(\ell)\}_R&=&\{b_\ga(k),b_\beta(\ell)\}_R=0\label{3.6a}\\
\{y_i(k),b_\ga(\ell)\}_R&=&\frac12\ga(h_i)b_\ga(k+\ell-N),\label{2stars}
\end{eqnarray}
for all $1\leq i,j\leq s$, $0\leq k,\ell\leq N$, and $\ga,\beta\in\Delta$.
\begin{theorem}\label{system-is-integrable-thm}
Let $f_{k\ell}:\ \mathcal{O}\rightarrow K$ be the functions $f_{k\ell}(L)=\tr_k(L^\ell)$ for all $L\in\mathcal{O}$, $0\leq k\leq N$, and $\ell\in\mathbb{Z}_+$.
\begin{enumerate}
\item[{\rm (i)}] The functions $f_{k\ell}$ commute with respect to the Poisson bracket $\{-,-\}_R$, for $0\leq k\leq N$ and $\ell\in\mathbb{Z}_+$.
\item[{\rm(ii)}] If $\fg$ is of type $D_n$ for $n\geq 4$, let $E(\fg)=\{1,3,5,\ldots, 2n-1\}$.  Otherwise, let $E(\fg)$ be the set of exponents\footnote{The {\em exponents} of $\fg$ are the positive integers $\ell$ for which the number of roots of height $\ell$ is strictly greater than the number of roots of height $\ell+1$.  The Poisson centre $S(\fg)^\fg$ is a polynomial algebra in algebraically independent homogeneoous generators of degree $m_1 +1,\ldots, m_s+1$, where $m_1,\ldots , m_s$ are the exponents of $\fg$.  See \cite[chapitre VIII \S8.3]{bourbaki-Lie7-8} for details.} of $\fg$.  Then the functions $f_{k\ell}$ are independent for $0\leq k\leq N$ and $\ell-1\in E(\fg)$.
\item[{\rm(iii)}] That is, $\mathbb{I}_N(\fg)=\{f_{k\ell}:\ 0\leq \ell\leq N,\ \ell-1\in E(\fg)\}$ is an integrable system on the orbit $\mathcal{O}$, generalizing the Toda systems for finite dimensional simple Lie algebras, with hamiltonian $H=\frac12\tr_N(L^2)$.
\end{enumerate}
\end{theorem}

\noindent
\proof\ \  (i)\  A straightforward calculation shows that the $r$-matrix (\ref{r-matrix}) controls the Poisson bracket between elements in the Lax matrix:
$$[r_{12}, L_1+L_2]=\{L_1,L_2\}_R,$$
where $L_1=L\ot 1$, $L_2=1\ot L$, $\{L_1,L_2\}_R=\sum_{i,j,k,\ell}\{a_{ij},a_{k\ell}\}_RE_{ij}\ot E_{k\ell},$ and $L=\sum_{i,j}a_{ij}E_{ij}\in\mathcal{O}$, viewed as an element of $\mathfrak{gl}((N+1)M).$  Expanding with the Leibniz rule, we have
\begin{eqnarray*}
\{L_1^m,L_2^n\}_R&=&\sum_{i=0}^{m-1}\sum_{j=0}^{n-1}L_1^{m-i-1}L_2^{n-j-1}\{L_1,L_2\}_RL_1^iL_2^j\\
&=&\sum_i\sum_j L_1^{m-i-1} L_2^{n-j-1}[r_{12}, L_1+L_2]L_1^i L_2^j\\
&=&\sum_i\sum_j[L_1^{m-i-1}L_2^{n-j-1}r_{12}L_1^i L_2^j, L_1+L_2],
\end{eqnarray*}
 using the fact that $L_1$ and $L_2$ commute.  The bilinear forms $(-|-)_k$ and $(-|-)_\ell$ introduced in Section 2 are symmetric, so
\begin{eqnarray*}
\tr_k\hbox{Tr}_{\ell,1}[X\ot Y, Z\ot W]&=&\tr_k\hbox{Tr}_{\ell,1}(XZ\ot YW-ZX\ot WY)\\
&=&\tr_k((X|Z)_\ell(YW-WY))\\
&=&(X|Z)_\ell((Y|W)_k-(W|Y)_k)\\
&=&0,
\end{eqnarray*}
for all $X,Y,Z,W\in\mathfrak{gl}((N+1)M)$.  In particular,
\begin{eqnarray*}
\{f_{km}(L),f_{\ell n}(L)\}_R&=&\{\tr_k(L^m),\tr_\ell(L^n)\}_R\\
&=&\tr_k\hbox{Tr}_{\ell,1}\{L_1^m,L_2^n\}_R\\
&=&\sum_i\sum_j\tr_k\hbox{Tr}_{\ell,1}[L_1^{m-i-1}L_2^{n-j-1}r_{12}L_1^i L_2^j, L_1+L_2]\\
&=&0,
\end{eqnarray*}
for all $L\in\mathcal{O}$, $0\leq k,\ell\leq N$, and $m,n\in\mathbb{Z}_+$.

\medskip

\noindent
(ii)\ The set $\{\tr(X^{\ell+1})\,:\,\ell\hbox{\ is an exponent of\ }\fg\}$ is an algebraically independent set of generators for the Poisson centre $S(\fg)^\fg$ outside of type $D_n$, $n\geq 4$, cf. \cite[Cor 2.4]{molev}.  The Poisson centre in type $D_n$ is generated by the trace functions $\tr(X^{\ell +1})$ for $\ell=1,3,\ldots, 2n-3$, together with the Pfaffian $\hbox{Pf}(n)$.  This set of polynomials is again algebraically independent, as is the set of trace functions $\{\tr(X^{\ell+1})\,:\,\ell=1,3,\ldots, 2n-1\}$, which generates the invariant subalgebra under the full orthogonal group:
\begin{eqnarray*}
K[\tr(X^{\ell+1})\,:\,\ell=1,3,\ldots, 2n-1]&=&K[\hbox{Pf}(n)^2,\,\tr(X^{\ell+1})\,:\,\ell=1,3,\ldots, 2n-3]\\
&=&S(\mathfrak{so}(2n))^{O(2n)}.
\end{eqnarray*}
The traditional proofs of algebraic independence use algebraic independence of the trace functions restricted to the Cartan subalgebra $\fh$.  Since $\fh\subset V=\hbox{Span}\,\mathcal{O}$, the polynomials $\tr(X^{\ell+1})$ are also algebraically independent as functions on $\mathcal{O}=\hbox{Ad}^*G_N^R\cdot x'$, for $\ell\in E(\fg)$.  In particular, this proves (ii) when $N=0$.

Now assume that the differentials of the functions $f_{k\ell}$ for $0\leq k\leq m$ and $\ell-1\in E(\fg)$ are linearly independent almost everywhere on $\mathcal{O}$, for all nonnegative integers $m$ strictly less than $N$.  Suppose there is a nontrivial dependence relation 
\begin{equation}\label{2etoiles}
\sum_{k,\ell}p_{k\ell}\,\hbox{d}f_{k\ell}=0
\end{equation}
at every point in a nonempty Euclidean open neighbourhood $U\subseteq V$, where $p_{k\ell}$ is a (not necessarily polynomial) function on $U$.  If $p_{N\ell}$ vanishes on $U$ for all $\ell$, then $\{\hbox{d}f_{k\ell}\,:\,0\leq k\leq N-1, \ell-1\in E(\fg)\}$ is linearly dependent on $U$, contradicting the induction hypothesis.

Without loss of generality, we can thus assume that the functions $p_{N\ell}$ are not all identically zero on $U$.  Note that
\begin{equation}\label{1csillag}
\hbox{d}f_{k\ell}(X)=\ell\,\tr_k(X^{\ell-1}\hbox{d}X),
\end{equation}
for each $X=X_0+X_1 u+\cdots+X_N u^N\in V$, and $\tr_k(X^{\ell-1}\hbox{d}X)$ can be expressed as a linear combination of terms of the form
\begin{equation}\label{2csillagok}
\tr(X_0^{\gs_0}\cdots X_N^{\gs_N}\hbox{d}X_j),
\end{equation}
where $\sum_{i=0}^N\gs_i=\ell-1$ and $\sum_{i=0}^Ni\gs_i=k-j$.  The matrix $X_N\in\fg$ is independent of the matrices $X_0,X_1,\ldots, X_{N-1}$ determining $\hbox{d}f_{k\ell}$ for $k<N$, so $X_N$ occurs in the expressions (\ref{2csillagok}) only when $k=N$.  In this case, (\ref{2csillagok}) simplifies to $\tr(X_0^{\ell-1}\hbox{d}X_N)$, and $\hbox{d}X_N$ is clearly linearly independent from $\hbox{d}X_0,\ldots,\hbox{d}X_{N-1}$.  By (\ref{2etoiles}) and (\ref{1csillag}), this means that the smaller expression
\begin{equation}\label{eqn4}
\sum_{\ell-1\in E(\fg)}\ell\,p_{N\ell}(X)\tr(X_0^{\ell-1}\hbox{d}X_N)=0,
\end{equation}
for all $X\in U$.  In terms of $(i,j)$-coordinates $x_{ij}$ of the matrix $X_N\in\fg\subseteq\mathfrak{gl}(M)$, the coefficients of the differentials $\hbox{d}x_{ij}$ are all zero in expression (\ref{eqn4}).  But then we also have
$$\sum_{\ell-1\in E(\fg)}p_{N\ell}(X)\hbox{d}f_{0\ell}=\sum_{\ell-1\in E(\fg)}\ell\, p_{N\ell}(X)\,\tr(X_0^{\ell-1}\hbox{d}X_0)=0$$
on $U$, since the component $X_0$ of $X$ is the only component contributing to the trace $\tr_0(X^{\ell-1}\hbox{d}X)$.  This contradicts the fact that $\{\hbox{d}f_{0\ell}\,:\,\ell-1\in E(\fg)\}$ is linearly independent almost everywhere on $V$, so (ii) holds by induction.

\medskip

\noindent
(iii)\ This is an immediate consequence of (i) and (ii).\qed

  

  \section{Equations of motion and their solution}
In this section, we derive equations of motion associated to the hamiltonian systems of Section 3, in terms of position and momentum coordinates $q_i(n)$, $p_i(n)$ on coadjoint orbits.  The equations can be solved by combining standard Lie group factorisation methods with new transformations called jet transformations.  These techniques are illustrated with explicit examples, including extensions of known soliton solutions for the rank $1$ infinite lattice.  As the material concerns time evolution of hamiltonian systems, we work over the field $\mathbb{R}$ of real numbers.


\subsection{Darboux coordinates and equations of motion}
  In the notation of Section 3, the Hamiltonian $H=\frac12\tr_N(L^2)$ is given by
\begin{equation}\label{hamiltonian}
  H=\frac12\sum_{\ell=0}^N\sum_{i=1}^s y_i(\ell)y_i(N-\ell)+\sum_{\ell=0}^N\sum_{\ga\in\Delta}b_\ga(\ell)b_\ga(N-\ell),
\end{equation}
where $L$ is the Lax matrix
\begin{equation}\label{lax-matrix}
  L=\sum_{\ell=0}^N\left(\sum_{i=1}^s y_i(\ell)h_i(\ell)+\sum_{\ga\in\Delta}b_\ga(\ell)(e_\ga(\ell)+f_\ga(\ell))\right)
  \end{equation}
parametrizing the $G_N^R$-coadjoint orbit $\mathcal{O}=\mathcal{O}_{x'}$.  We now introduce a system of (global) position and momentum coordinates $q_i(n),p_i(n)$ on $\mathcal{O}$ in terms of the coefficient functions $y_i(\ell),b_\ga(\ell)$.
\begin{proposition}\label{darboux-defn-lemma} Let $\mathcal{F}(\mathcal{O})$ be the algebra of (real-valued) smooth functions on $\mathcal{O}$.  The equations
  $$\sum_{\ell=0}^Nb_\ga(\ell)v^\ell=\exp\left(\frac12\sum_{n=0}^N\sum_{i=1}^s\ga(h_i)q_i(n)v^n\right)\in\mathcal{F}(\mathcal{O})\ot_{\mathbb{R}}\mathbb{R}[v]/\langle v^{N+1}\rangle,$$
  for $\ga\in\Delta$, uniquely determine a family of functions $q_i(n)\in\mathcal{F}(\mathcal{O})$.
\end{proposition}

\bigskip

\noindent
\proof\ \ Let $\ga$ be a simple root, and define $Q_n=Q(\ga,n)=\sum_{i=1}^s\ga(h_i)q_i(n)$.  Writing
$$\left(\sum_{n=0}^Nc_nv^n\right)_\ell$$
for the coefficient $c_\ell$ of $v^\ell$ in any formal series $\sum_{n=0}^Nc_nv^n\in\mathcal{F}(\mathcal{O})\ot_\mathbb{R}\mathbb{R}[v]/\langle v^{N+1}\rangle$, we see that
\begin{align}\nonumber
  b_\ga(\ell)=&\left(\exp\sum_{n=0}^N\frac{Q_n}{2}v^n\right)_\ell=\left(\prod_{n=0}^N\exp(\frac{Q_n}{2}v^n)\right)_\ell\\
  =&e^{Q_0/2}\sum_{\gs\vdash\ell}\frac{Q_1^{\gs_1}Q_2^{\gs_2}\cdots Q_N^{\gs_N}}{2^{|\gs|}\gs_1!\gs_2!\cdots\gs_N!}=e^{Q_0/2}\sum_{\gs\vdash\ell}\frac{Q_1^{\gs_1}Q_2^{\gs_2}\cdots Q_\ell^{\gs_\ell}}{2^{|\gs|}\gs_1!\gs_2!\cdots\gs_\ell!},
  \label{b(l)-formula}
  \end{align}
where $|\gs|=\gs_1+\cdots+\gs_N$, and the final two sums are taken over all partitions $\gs=(\gs_1,\ldots,\gs_N)\in\mathbb{Z}_+^N$ with $\sum_{i=1}^Ni\gs_i=\ell$.  In particular,
$$b_\ga(0)=e^{Q_0/2}=\exp\left(\frac12\sum_{i=1}^s\ga(h_i)q_i(0)\right),$$
and for the base $\Delta$ of simple roots $\ga_1,\ldots,\ga_s$, we have
\begin{eqnarray}\left(\begin{array}{cccc}\ga_1(h_1) & \ga_1(h_2) & \cdots & \ga_1(h_s)\\
    \ga_2(h_1) & \ga_2(h_2) & \cdots & \ga_2(h_s)\\
    \vdots & \vdots & \ddots & \vdots \\
    \ga_s(h_1) & \ga_s(h_2) & \cdots &\ga_s(h_s)
  \end{array}\right)\left(\begin{array}{c}q_1(0)\\ q_2(0)\\ \vdots \\ q_s(0)\end{array}\right)=\left(\begin{array}{c}2\log b_{\ga_1}(0)\\2\log b_{\ga_2}(0)\\\vdots\\2\log b_{\ga_s}(0)\end{array}\right).\nonumber
\end{eqnarray}
  Note that the coefficient functions $b_{\ga_i}(0)$ are positive everywhere on $\mathcal{O}$, as can be seen by exponentiating Formulas (\ref{coadjoint-formulas}) for the $R$-coadjoint action.  The matrix $(\ga_i(h_j))_{ij}$ is nonsingular, since $\{h_1,\ldots,h_s\}$ is a basis of $\mathfrak{h}$ and $\{\ga_1,\ldots, \ga_s\}$ is a basis of $\fh^*$.  The functions $q_i(0)$ are thus uniquely determined.

  We now proceed by induction, assuming that $q_i(k)\in\mathcal{F}(\mathcal{O})$ is uniquely determined for all $k<n$.  By (\ref{b(l)-formula}), 
  $$b_\ga(n)=e^{Q_0/2}\sum_{\gs\vdash n}\frac{Q_1^{\gs_1}Q_2^{\gs_2}\cdots Q_n^{\gs_n}}{2^{|\gs|}\gs_1!\gs_2!\cdots\gs_n!},$$
  where $Q_0,\ldots,Q_{n-1}$ depend only on $\ga$ and $q_i(k)$ for $1\leq i\leq s$ and $0\leq k<n$.  The only summand depending on $q_i(n)$ is thus the term corresponding to the partition with $\gs_1=\gs_2=\cdots=\gs_{n-1}=0$ and $\gs_n=1$.  That is,
  $$b_\ga(n)=c_\ga+d_\ga Q_n,$$
  for some functions $c_\ga,d_\ga$ depending only on $\ga$ and $q_i(k)$ for $1\leq i\leq s$ and $0\leq k<n$.  Note that $d_\ga=\frac12 e^{Q_0/2}$ is never zero.  By varying the simple root $\ga$, we now have another system of $s$ linear equations
  $$\sum_{i=1}^s\ga_j(h_i)q_i(n)=\frac{b_{\ga_j}(n)-c_{\ga_j}}{d_{\ga_j}}$$
  in $s$ unknowns $q_1(n),\ldots,q_s(n)$, with a unique solution since the matrix $(\ga_j(h_i))_{ij}$ is nonsingular.\qed

  \bigskip

  \begin{proposition} \label{darboux-bracket-prop}Let $p_i(n)=y_i(n)$, and define $q_i(n)$ as in Proposition \ref{darboux-defn-lemma}.  The set $S=\{p_i(n),q_i(n)\ :\ 1\leq i\leq s,\ 0\leq n\leq N\}$ is then a system of independent generators of the Poisson algebra $(\mathcal{F}(\mathcal{O}_{x'}),\{-,-\}_R)$ with brackets
    \begin{align}
      \{p_i(m),p_j(n)\}_R&=\{q_i(m),q_j(n)\}_R=0 \label{csillag}\\
      \{p_i(m),q_j(n)\}_R&=\delta_{ij}\delta_{m+n,N},\label{csillagok}
    \end{align}
    for all $1\leq i,j\leq s$ and $0\leq m,n\leq N$.
\end{proposition}

  \bigskip

  \noindent
  \proof\ \ By Proposition \ref{coadjoint-orbit-description-proposition}, the coadjoint orbit $\mathcal{O}=\mathcal{O}_{x'}$ is an open subset of a $2s(N+1)$-dimensional vector space $V$.  The algebra $\mathcal{F}(\mathcal{O})$ is thus generated by the $2s(N+1)$ coordinate functions $y_i(n),b_\ga(n)$ on $\mathcal{O}$.  Since these coordinate functions are also in the subalgebra generated by $S$, the set $S$ clearly generates $\mathcal{F}(\mathcal{O})$. Its elements are necessarily independent (i.e.~have linearly independent differentials), since the number of elements in $S$ is exactly the dimension of $V$.

  The Poisson brackets of members of any generating set of $\mathcal{F}(\mathcal{O})$ obviously determine the Poisson brackets on any other generating set, so to complete the proof, it is enough to check that, if the members of $S$ satisfy (\ref{csillag}) and (\ref{csillagok}), then they satisfy the following relations imposed by Equations (\ref{3.6a}) and (\ref{2stars}):
  \begin{equation} \{p_i(m),p_j(n)\}_R=0\label{darboux-pp},
  \end{equation}
  \begin{equation}\label{darboux-qq}
    \left\{\left(\exp\frac12\sum_{\ell=0}^N\sum_{k=1}^s\ga(h_k)q_k(\ell)v^\ell\right)_m,\left(\exp\frac12\sum_{\ell=0}^N\sum_{k=1}^s\beta(h_k)q_k(\ell)v^\ell\right)_n\right\}_R=0,\end{equation}
    \begin{align} \left\{p_i(m),\left(\exp\frac12\sum_{\ell=0}^N\right.\right.&\left.\left.\sum_{k=1}^s\ga(h_k)q_k(\ell)v^\ell\right)_n\right\}_R\nonumber\\
      & =\frac12\ga(h_i)\left(\exp\frac12\sum_{\ell=0}^N\sum_{k=1}^s\ga(h_k)q_k(\ell)v^\ell\right)_{n+m-N}.\label{darboux-pq}
    \end{align}
    The relations (\ref{darboux-pp}) and (\ref{darboux-qq}) are obvious from (\ref{csillag}).  To see that (\ref{darboux-pq}) also holds, note that if $S$ satisfies (\ref{csillag}) and (\ref{csillagok}), then $\{p_i(m),-\}_R$ acts by $\frac{\partial}{\partial q_i(N-m)}$ on functions of the position variables $q_j(\ell)$, so
    \begin{align} \left\{p_i(m),\left(\exp\frac12\sum_{\ell=0}^N\right.\right.&\left.\left.\sum_{k=1}^s\ga(h_k)q_k(\ell)v^\ell\right)_n\right\}_R\nonumber\\
      &=\left(\frac{\partial}{\partial q_i(N-m)}\exp\frac12\sum_{\ell=0}^N\sum_{k=1}^s\ga(h_k)q_k(\ell)v^\ell\right)_n\nonumber\\
      &=\frac12\ga(h_i)\left(\exp\frac12\sum_{\ell=0}^N\sum_{k=1}^s\ga(h_k)q_k(\ell)v^\ell\right)_{n+m-N}\nonumber .
    \end{align}
  \qed

  \bigskip

  The expression (\ref{hamiltonian}) for the Hamiltonian can be rewritten as
  $$H=\frac12\sum_{i=1}^s(y_i(v)^2)_N+\sum_{\ga\in\Delta}(b_{\ga}(v)^2)_N,$$
  where $y_i(v)$ and $b_\ga(v)$ are the generating functions
  $$y_i(v)=\sum_{n=0}^Ny_i(n)v^n\hbox{\ \ and \ \ }b_\ga(v)=\sum_{n=0}^Nb_\ga(n)v^n,$$
respectively,  in $\mathcal{F}(\mathcal{O})\ot\mathbb{R}[v]/\langle v^{N+1}\rangle$.  In terms of the Darboux coordinates $\{p_i(n),q_i(n)\}$ introduced in Propositions \ref{darboux-defn-lemma} and \ref{darboux-bracket-prop}, the Hamiltonian becomes:
  \begin{align*}
    H&=\frac12\sum_{i=1}^s\sum_{j=0}^Np_i(j)p_i(N-j)+\sum_{\ga\in\Delta}\left(\exp\sum_{n=0}^N\sum_{i=1}^s\ga(h_i)q_i(n)v^n\right)_N\\
    &=\frac12\sum_{i=1}^s\sum_{j=0}^Np_i(j)p_i(N-j)+\sum_{\ga\in\Delta}e^{Q(\ga,0)}\sum_{\gs\vdash N}\frac{Q_\ga^\gs}{\gs!},
    \end{align*}
  where we use multi-index notation $Q_\ga^\gs=Q(\ga,1)^{\gs_1}\cdots Q(\ga,N)^{\gs_N}$ and $\gs!=\gs_1!\cdots\gs_N!$ for partitions $\gs=(\gs_1,\ldots,\gs_N)\in\mathbb{Z}_+^N$ of the form $\sum_{i=1}^N i\gs_i=N$, with $Q(\ga,n)$ defined as $\sum_{i=1}^s\ga(h_i)q_i(n)$, as in the proof of Proposition \ref{darboux-defn-lemma}.

  Calculating directly, and writing $\bigdot{p_i}(n)$ and $\bigdot{q_i}(n)$ for the time derivatives $\frac{dp_i(n)}{dt}$ and $\frac{dq_i(n)}{dt}$, the Hamiltonian system $(\mathcal{O},H)$ has equations of motion
  \begin{align}
    \bigdot{q_i}(n)&=\{H,q_i(n)\}_R=\left\{\frac12\sum_{\ell=0}^N\sum_{k=1}^sp_k(\ell)p_k(N-\ell),q_i(n)\right\}_R=p_i(n)\label{q_i-dot}\\
    \bigdot{p_i}(n)&=\{H,p_i(n)\}_R \nonumber\\
    &=\big(\big\{\sum_{\ga\in\Delta}b_{\ga}(v)^2,y_i(n)\big\}_R\big)_N\nonumber\\
    &=-\sum_{\ga\in\Delta}\ga(h_i)(b_\ga(v)^2)_n\nonumber\\
&=-\sum_{\ga\in\Delta}\ga(h_i)\left(\exp\sum_{k=0}^NQ(\ga,k)v^k\right)_n\nonumber\\
    &=-\sum_{\ga\in\Delta}\left(\ga(h_i)e^{Q(\ga,0)}\sum_{\gs\vdash n}\frac{Q_\ga^\gs}{\gs!}\right)\nonumber\\
      &=-\sum_{\ga\in\Delta}\left(\ga(h_i)\exp\left(\sum_{j=1}^s\ga(h_j)q_j(0)\right)\sum_{\gs\vdash n}\prod_{\ell=1}^n\frac{1}{\gs_{\ell}!}\left(\sum_{k=1}^s\ga(h_k)q_k(\ell)\right)^{\gs_\ell}\right)\label{p_i-dot}
    \end{align}
  for $1\leq i\leq s$ and $0\leq n\leq N$.

  For example, if $N=0$, we recover the classical open Toda evolution equations introduced by Toda in Type A, and their extensions to other root systems \cite{toda67,bogoyavlensky76,kostant79}:
  \begin{eqnarray}
    \bigdot{q_i}(0)&=&p_i(0)\label{N=0-eqns-of-motion-1}\\
    \bigdot{p_i}(0)&=&-\sum_{\ga\in\Delta}\big(\ga(h_i)\exp\sum_{j=1}^s\ga(h_j)q_j(0)\big).\label{N=0-eqns-of-motion-2}
  \end{eqnarray}
  When $N=2$, we have a system of $3s$ particles in a $6s$-dimensional phase space:\begin{eqnarray*}
    \bigdot{q_i}(n)&=&p_i(n),\hbox{\ for\ }n=0,1,2\\
    \bigdot{p_i}(0)&=&-\sum_{\ga\in\Delta}\big(\ga(h_i)\exp\sum_{j=1}^s\ga(h_j)q_j(0)\big)\\
    \bigdot{p_i}(1)&=&-\sum_{\ga\in\Delta}\big(\ga(h_i)\exp\big(\sum_{j=1}^s\ga(h_j)q_j(0)\big)\sum_{k=1}^s\ga(h_k)q_k(1)\big)\\
    \bigdot{p_i}(2)&=&-\sum_{\ga\in\Delta}\left(\ga(h_i)\exp\big(\sum_{j=1}^s\ga(h_j)q_j(0)\big)\big(\frac12\big(\sum_{k=1}^s\ga(h_k)q_k(1)\big)^2+\sum_{i=1}^s\ga(h_i)q_i(2)\big)\right).
  \end{eqnarray*}

The equations of motion (\ref{q_i-dot}) and (\ref{p_i-dot}) remain well defined in more general settings, even those without access to Liouville integrable systems and hamiltonian functions.  All that is needed is an appropriate choice of root system and Cartan elements $h_j$ to determine the expression in the last line of (\ref{p_i-dot}).  For example, Toda \cite{toda67} studied equations associated to an infinite lattice:
\begin{eqnarray}
\bigdot{q_i}&=&p_i\nonumber\\
\bigdot{p_i}&=&-e^{q_i-q_{i+1}}+e^{q_{i-1}-q_i},\hbox{\ \ for\ }i\in\mathbb{Z}.\label{Toda-Ainfty}
\end{eqnarray}
These equations correspond to the locally finite Lie algebra $\mathfrak{gl}(\infty)$ of doubly infinite matrices with only finitely many nonzero entries, where $\{h_j\,:\,j\in\mathbb{Z}\}$ is an orthonormal basis of the diagonal matrices in $\mathfrak{gl}(\infty)$, and $\Delta=\{\epsilon_i-\epsilon_{i+1}\,:\,i\in\mathbb{Z}\}$ is a set of simple roots with $\epsilon_i(h_j)=\delta_{ij}$.  Inserting this data into (\ref{p_i-dot}) gives equations for Takiff algebras of type $A_\infty$, whose solution will be discussed in Example \ref{infinite-Toda-example}:
\begin{eqnarray}
\bigdot{q_i}(n)&=&p_i(n)\nonumber\\
\bigdot{p_i}(n)&=&-\exp\big(q_i(0)-q_{i+1}(0)\big)\sum_{\gs\vdash n}\prod_{\ell=1}^n\frac{1}{\gs_\ell!}\big(q_i(\ell)-q_{i+1}(\ell)\big)^{\gs_\ell}\nonumber\\
  &\ &\quad\quad +\exp\big(q_{i-1}(0)-q_i(0)\big)\sum_{\gs\vdash n}\prod_{\ell=1}^n\frac{1}{\gs_\ell!}\big(q_{i-1}(\ell)-q_i(\ell)\big)^{\gs_\ell}.\label{Takiff-Ainfty}
\end{eqnarray}

  \subsection{Exact solutions by factorisation}\label{section-exact-solutions-by-factorisation}
  The equations of motion (\ref{q_i-dot}) and (\ref{p_i-dot}) are equivalent to a Lax presentation
  \begin{equation}\label{Lax-presentation}
\bigdot{L}=[M,L],
    \end{equation}
  where $L=L(t)$ is the Lax matrix (\ref{lax-matrix}) and
  $$M=\frac12\sum_{j=0}^N\sum_{\ga\in\Delta}b_\ga(j)(e_\ga(j)-f_\ga(j)).$$
  This can be verified directly by changing the matrix entries in (\ref{Lax-presentation}) to the Darboux coordinates introduced in Proposition \ref{darboux-bracket-prop} and comparing matrix entries on the two sides of Equation (\ref{Lax-presentation}).  The Lax framework lets us exactly solve the Hamiltonian system using factorisation in Lie groups associated to truncated current algebras.

  The strategy consists of expressing the exponential
  $$\exp(-t\,dH(L(0))=\exp(-t\,L(0))$$
  as a product of certain integral curves $\theta_-(t)^{-1}$ and $\theta_+(t)$ on Lie subgroups of $G_N$, where $L(0)$ is the Lax matrix at time $t=0$ and ${d}H$ is the differential of the Hamiltonian function $H\in\mathcal{F}(\mathcal{O})$.  The adjoint action of $\theta_+(t)$ then gives solutions to the equations of motion by comparing entries of the matrices in the expression  $$L(t)=\theta_+(t)L(0)\theta_+^{-1}(t).$$

  We illustrate this method with exact solutions for the extremal case (minimal root lattice) when $\fg=\mathfrak{sl}_2(\mathbb{R})$, deriving solutions in terms of Darboux coordinates for the $3$-body problem associated with $N=2$, for certain initial conditions.  To simplify notation, we write $q_n$ for the (rescaled) position variable $\frac{1}{\sqrt{2}}q_1(n)$ for $n=0,1,\ldots,N$.  Similarly, $p_n$ will be the momentum $\frac{1}{\sqrt{2}}p_1(n)$, where all masses are taken to be $1$.  When we wish to emphasize evolution in terms of time $t$, we write $q_n(t)$ and $p_n(t)$ for $q_n$ and $p_n$, respectively.  Note that in this section, $q_n(0)$ and $p_n(0)$ will denote the initial position and initial velocity of the $n$th particle.

  In the case of $\fg=\mathfrak{sl}_2(\mathbb{R})$, the Lax matrix $L=L(t)$ may be viewed as the $2\times 2$ matrix
  \begin{equation}\label{lax-matrix-sl2}
    L(t)=\left(\begin{array}{cc}
      y(v,t) & b(v,t)\\
      b(v,t) & -y(v,t)
    \end{array}\right),\end{equation}
  where $y(v,t)=\sum_{n=0}^Np_n(t)v^n$ and $b(v,t)=\exp\sum_{n=0}^Nq_n(t)v^n$, in which the formal variable $v$ is identified with its image in $\mathbb{R}[v]/\langle v^{N+1}\rangle$.

  Diagonalizing the matrix $-t\,L(0)$, we see that
  $$P\left(\begin{array}{cc}
    -ty & -tb\\
    -tb & ty
    \end{array}\right)P^{-1}=\left(\begin{array}{cc}
    -t\beta & 0\\
    0 & t\beta
  \end{array}\right),$$
  where $y=y(v,0)$, $b=b(v,0)$, $\beta$ is a square root of $y^2+b^2$,
  $$P=\left(\begin{array}{cc}
    y+\beta & b\\
    y-\beta & b
    \end{array}\right)\hbox{\ \ and \ \ }P^{-1}=\frac{1}{2b\beta}\left(\begin{array}{cc}
    b & -b\\
    \beta-y & y+\beta
  \end{array}\right),$$
  noting that $\beta$, $\frac{1}{\beta}$, and $\frac{1}{2b\beta}$ always exist by Lemma \ref{invertability-and-sqrt}, since the coefficients of $v^0$ in the expressions for $y^2+b^2$ and $b$ are always positive, and the coefficients of $v^0$ in $\beta$ and $2b\beta$ are thus always nonzero. Without loss of generality, we choose $\beta$ so that its constant term is the positive square root $\sqrt{p_0(0)^2+e^{2q_0(0)}}$.  Therefore,
  \begin{align*}
    \exp(-t\,L(0))&=\exp\left(P^{-1}\left(\begin{array}{cc}
    -t\beta & 0\\
    0 & t\beta
    \end{array}\right)P\right)\\
&=P^{-1}\left(\begin{array}{ll}
    e^{-t\beta} & 0\\
    0 & e^{t\beta}
    \end{array}\right)P\\
    &=\frac{1}{\beta}\left(\begin{array}{cc}
    \beta\cosh{t\beta}-y\sinh{t\beta} & -b\sinh{t\beta}\\
    -b\sinh{t\beta} & \beta\cosh{t\beta}+y\sinh{t\beta}
  \end{array}\right),
  \end{align*}
  where the hyperbolic trigonometric functions are defined as usual, with all computations taking place in $\mathbb{R}[v]/\langle v^{N+1}\rangle$:
  \begin{align*}
    \sinh{t\beta}&=\frac12(e^{t\beta}-e^{-t\beta})\\
    \cosh{t\beta}&=\frac12(e^{t\beta}+e^{-t\beta}).
    \end{align*}
  Let $A=\cosh{t\beta}-\frac{y}{\beta}\sinh{t\beta}$, $B=\frac{-b}{\beta}\sinh{t\beta}$, and $C=\cosh{t\beta}+\frac{y}{\beta}\sinh{t\beta}.$  Then $\sqrt{A}$ and $\frac{1}{\sqrt{A}}$ exist by Lemma \ref{invertability-and-sqrt}, so
  $$\exp(-t\,L(0))=\left(\begin{array}{cc}A & B\\ B & C\end{array}\right)=\theta_-(t)^{-1}\theta_+(t),$$
    where
    $$\theta_+(t)=\left(\begin{array}{cc}\sqrt{A} & B/\sqrt{A}\\ 0 & 1/\sqrt{A}\end{array}\right)\hbox{\ \ and\ \ }\theta_-(t)=\left(\begin{array}{rr}1/\sqrt{A} & 0\\ -B/\sqrt{A} & \sqrt{A}\end{array}\right).$$
    The curves $\theta_+(t)$ and $\theta_-(t)$ belong to the Lie subgroups corresponding to the standard and opposite Borels of $G_N=SL_2(\mathbb{R}[u]/\langle u^{N+1}\rangle)$, respectively.  Their Cartan components
    $$\left(\begin{array}{cc}\sqrt{A} & 0\\ 0 & 1/\sqrt{A}\end{array}\right)\hbox{\ \ and \ \ }\left(\begin{array}{cc}1/\sqrt{A} & 0\\ 0 & \sqrt{A}\end{array}\right)$$
    are inverse to one another.  By general principles of the Lax formalism, conjugation of $L(0)$ by $\theta_+(t)$ now gives a Lax expression for solutions to the equations of motion
    $$L(t)=\theta_+(t)L(0)\theta_+(t)^{-1}.$$  Calculating directly, we see that
    
    $$\left(\begin{array}{rr}y(v,t) & b(v,t)\\ b(v,t) & -y(v,t)\end{array}\right)=\frac{1}{A}\left(\begin{array}{cc}Ay+Bb & b\\ b & -Ay-Bb\end{array}\right),$$
      so
      \begin{align}
        y(v,t)&=y-\frac{b^2\sinh{t\beta}}{\beta\cosh{t\beta}-y\sinh{t\beta}}\label{Lax-solution1}\\
        b(v,t)&=\frac{b\beta}{\beta\cosh{t\beta}-y\sinh{t\beta}}.\label{Lax-solution2}   \end{align}
      Equations (\ref{Lax-solution1}) and (\ref{Lax-solution2}) give solutions to the equations of motion (\ref{q_i-dot}) and (\ref{p_i-dot}) when $\fg=\mathfrak{sl}_2(\mathbb{R})$ and allow explicit calculation of the Darboux coordinates $q_n(t),p_n(t)$ with respect to any family of initial conditions $\{q_n(0),p_n(0)\}_{0\leq n\leq N}$.  This calculation is straightforward, but rather onerous for all but the smallest values of $n$.

      We illustrate the process for the $3$-body problem associated with $\fg=\mathfrak{sl}_2(\mathbb{R})$ and $N=2$, for initial velocities $p_0(0)=p_1(0)=p_2(0)=0$ and arbitrary initial positions $q_0(0),q_1(0),$ and $q_2(0)$.  In this case, the Hamiltonian is
      $$H=(p_1^2+2p_0p_2)+2(q_1^2+q_2)e^{2q_0},$$
      giving equations of motion
      \begin{align}
        \bigdot{q}_n&=p_n\hbox{\ \ for\ }n=0,1,2 \nonumber\\
        \bigdot{p}_0&=-e^{2q_0} \nonumber\\
        \bigdot{p}_1&=-2q_1e^{2q_0}\nonumber \\
        \bigdot{p}_2&=-2(q_1^2+q_2)e^{2q_0},\label{eqns-of-motion-N=2}
        \end{align}
      keeping in mind that $\{q_m,q_n\}_R=\{p_m,p_n\}_R=0$ and $\{p_m,q_n\}_R=\frac12\delta_{m+n,2}$.  Under our initial conditions, Equations (\ref{Lax-solution1}) and (\ref{Lax-solution2}) simplify to
      \begin{align*}
        y(v,t)&=b\tanh{tb}\\
        b(v,t)&=b\,\hbox{sech}\,tb.
        \end{align*}
      Expanding in Darboux coordinates,
      $$b(v,t)=\exp\sum_{n=0}^2q_n(t)v^n=e^{q_0(t)}+q_1(t)e^{q_0(t)}v+\frac12(q_1(t)^2+2q_2(t))e^{q_0(0)}v^2$$
      and
      \begin{align*}
        \hbox{sech}\,tb&=\sech\big(te^{q_0(0)}\big)-tq_1(0)e^{q_0(0)}\sech\big(te^{q_0(0)}\big)\tanh\big(te^{q_0(0)}\big)v\\
        &\ \ \ +\frac12\left(t^2q_1(0)^2e^{2q_0(0)}\big(2\tanh^2(te^{q_0(0)})-1\big)\right.\\&\ \ \ - \left. t\left(q_1(0)^2+2q_2(0)\right)e^{q_0(0)}\tanh\big(te^{q_0(0)}\big)\right)\sech\big(te^{q_0}\big)v^2,
        \end{align*}
      from which it easily follows that
      \begin{align}
        q_0(t)&=q_0(0)-\log(\cosh(te^{q_0(0)}))\nonumber\\
        q_1(t)&=q_1(0)-tq_1(0)e^{q_0(0)}\tanh(te^{q_0(0)})\nonumber\\
        q_2(t)&=q_2(0)-\frac{t}{2}(q_1(0)^2+2q_2(0))e^{q_0(0)}\tanh(te^{q_0(0)})-\frac{t^2}{2}q_1(0)^2e^{2q_0(0)}\sech^2(te^{q_0(0)})\label{N=2-solution}
        \end{align}
is a (global) solution to the equations of motion (\ref{eqns-of-motion-N=2}).

  \subsection{Jet transformations and general solutions}\label{jet-transformations-subsection}
The standard Lie group factorisation methods described in Section \ref{section-exact-solutions-by-factorisation} are elegant but computationally unwieldy in the setting of Takiff algebras $\fg_N$, without imposing particular initial conditions and restricting to small values of $N$, as illustrated in the example above.  In this section, we introduce differential operators called {\em jet transformations} to generate solutions for the equations of motion attached to Takiff algebras for arbitrary $N$, whenever there are known solutions available for $\fg$.  As solutions for $\fg$ can always be obtained \cite{kostant79}, jet transformations provide solutions for all Takiff algebras $\fg_N$.

Specializing to type $A_1$, we obtain general solutions for $N=2$, recovering (\ref{N=2-solution}) after imposing the same initial conditions.  Applied to the rank $1$ infinite lattice Toda model, jet transformations extend KdV soliton solutions to the larger Takiff systems.

Let $\fg$ be a split simple, affine, or locally finite Lie algebra, with root system $\Phi$ and base $\Delta$ of simple roots $\ga_i$, $i\in I$.  Let $x_{in},z_{in}\in\mathbb{R}$ for $i\in I$ and $n>0$, and let $\mathcal{C}=\mathcal{C}^\infty(t,x_{i0},z_{i0}\,:\,i\in I)$ be the associative $\mathbb{R}$-algebra of smooth functions in independent real variables $t$, $x_{i0}$, and $z_{i0}$, for $i\in I$.  Let $\delta:\ \mathcal{C}_N\rightarrow\mathcal{C}_N$ be the derivation
$$\delta=\sum_{n=0}^N\sum_{i\in I}\left(x_{in}\frac{\partial}{\partial x_{i0}}+z_{in}\frac{\partial}{\partial z_{i0}}\right)\ot v^n,$$
where $\mathcal{C}_N=\mathcal{C}\ot_\mathbb{R}\mathbb{R}[v]/\langle v^{N+1}\rangle$ is the associative algebra of truncated currents, for some $N\geq 0$.  When there is no confusion, we will regard $\mathcal{C}=\mathcal{C}\ot_\mathbb{R}\mathbb{R}$ as a subalgebra of $\mathcal{C}_N$ and suppress the tensor symbol from elements and transformations of $\mathcal{C}_N$.  For $n=0,\ldots, N$, the linear transformation
$$D_n=(\exp\delta)_n\ :\ \mathcal{C}\rightarrow\mathcal{C}$$
will be called the {\em jet transformation} of order $n$.  Explicitly,
\begin{equation}\label{D_n}
D_n=\sum_{\gs\vdash n}\prod_{\ell=1}^n\frac{1}{\gs_\ell !}\left(\sum_{i\in I}x_{i\ell}\frac{\partial}{\partial x_{i0}}+z_{i\ell}\frac{\partial}{\partial z_{i0}}\right)^{\gs_\ell}.
\end{equation}
\begin{theorem}\label{jet-theorem}
Let $q_j(0)(t),p_j(0)(t)\in\mathcal{C}$ be a solution of the equations of motion (\ref{N=0-eqns-of-motion-1}) and (\ref{N=0-eqns-of-motion-2}) of the Toda system for $\fg$, satisfying initial conditions $q_j(0)(0)=x_{j0}$ and $p_j(0)(0)=z_{j0}$ for all $j\in I$.  Then
\begin{eqnarray*}
q_j(n)(t)&=&D_n(q_j(0)(t))\\
p_j(n)(t)&=&D_n(p_j(0)(t)),\hbox{\ for\ }n=0,\ldots, N\hbox{\ and\ }j\in I,
\end{eqnarray*}
is a solution of the equations of motion (\ref{q_i-dot}) and (\ref{p_i-dot}) for $\fg_N$, for all $N\in\mathbb{Z}_+$.  These solutions satisfy the initial conditions
\begin{eqnarray*}
q_j(n)(0)&=&x_{jn}\\
p_j(n)(0)&=&z_{jn}
\end{eqnarray*}
for all $n=0,\ldots, N$ and $j\in I$.
\end{theorem}
\proof\ \  Let $q_j(n)=q_j(n)(t)=D_n(q_j(0)(t))$ and $p_j(n)=p_j(n)(t)=D_n(p_j(0)(t))$ for all $n=0,\ldots, N$ and $j\in I$.  As $t$ and $q_i(0)(0)$ are independent for all $i$, the derivation $\delta$ commutes with $\frac{\partial}{\partial t}\ot 1$ on $\mathcal{C}_N$, so $\frac{\partial}{\partial t}$ commutes with $D_n=(\exp\delta)_n$ on $\mathcal{C}$.  In particular,
\begin{eqnarray*}
\bigdot{q}_j(n)&=&\frac{\partial}{\partial t}D_n(q_j(0))\\
&=&D_n(\bigdot{q}_j(0))\\
&=&D_n(p_j(0))\\
&=&p_j(n)
\end{eqnarray*}
since $\bigdot{q}_j(0)=p_j(0)$, by (\ref{N=0-eqns-of-motion-1}).  Moreover,
\begin{eqnarray*}
\sum_{n=0}^N \bigddot{q}_j(n)v^n&=&\sum_{n=0}^N\frac{\partial^2}{\partial t^2}D_n(q_j(0))v^n\\
&=&\sum_{n=0}^N D_n(\bigddot{q}_j(0))v^n,
\end{eqnarray*}
and
$$\bigddot{q}_j(0)=-\sum_{\ga\in\Delta}\ga(h_j)\exp Q(\ga,0),$$
by (\ref{N=0-eqns-of-motion-1}) and (\ref{N=0-eqns-of-motion-2}),
where $Q(\ga,n)=\sum_{i\in I}\ga(h_i)q_i(n)$ for all $\ga\in\Delta$ and $n=0,\ldots, N$.  But then
\begin{eqnarray*}
\sum_{n=0}^N\bigddot{q}_j(n)v^n&=&-\sum_{n=0}^N D_n\left(\sum_{\ga\in\Delta}\ga(h_j)\exp Q(\ga,0)\right)v^n\\
&=&-\sum_{\ga\in\Delta}\ga(h_j)\sum_{n=0}^N D_n\big(\exp Q(\ga,0)\big)v^n\\
&=&-\sum_{\ga\in\Delta}\ga(h_j)(\exp\delta)(\exp Q(\ga,0)).
\end{eqnarray*}

Since $\delta$ is a derivation, its exponential
$$\exp\delta=\sum_{n=0}^N D_nv^n\ :\ \mathcal{C}_N\rightarrow\mathcal{C}_N$$
is an automorphism, so $\exp\delta$ commutes with the exponential map.  That is,
\begin{eqnarray*}
(\exp\delta)(\exp Q(\ga,0))&=&\exp\big((\exp\delta)Q(\ga,0)\big)\\
&=&\exp\sum_{n=0}^ND_nQ(\ga,0)v^n\\
&=&\exp\sum_{n=0}^NQ(\ga,n)v^n,
\end{eqnarray*}
and
$$\sum_{n=0}^N\bigddot{q}_j(n)v^n=-\sum_{\ga\in\Delta}\ga(h_j)\exp\sum_{k=0}^NQ(\ga,k)v^k.$$
Comparing coefficients of $v^n$ now shows that $q_j(n)$ satisfies the $\fg_N$ equations of motion:
\begin{eqnarray*}
\bigdot{q}_j(n)&=&p_j(n)\\
\bigdot{p}_j(n)&=&-\sum_{\ga\in\Delta}\ga(h_j)\left(\exp\sum_{k=0}^NQ(\ga,k)v^k\right)_n\\
&=&-\sum_{\ga\in\Delta}\left(\ga(h_j)\exp\left(\sum_{i=1}^s\ga(h_i)q_i(0)\right)\sum_{\gs\vdash n}\prod_{\ell=1}^n\frac{1}{\gs_{\ell}!}\left(\sum_{k=1}^s\ga(h_k)q_k(\ell)\right)^{\gs_\ell}\right),
\end{eqnarray*}
 for $n=0,\ldots, N$, $j\in I$, and all $N>0$.

To see that $q_j(n)$ and $p_j(n)$ satisfy the specified initial conditions, note that $x_{jk}$ and $z_{jk}$ are independent of $t$, so the evaluation map $\ev:\ \mathcal{C}\rightarrow\mathcal{C}$, evaluating the variable $t$ at the value $t=0$, commutes with $D_n$.  That is, 
\begin{eqnarray*}
q_j(n)(0)&=&\ev\,q_j(n)(t)\\
&=&\ev\,D_n(q_j(0)(t))\\
&=&D_n\,\ev\,q_j(0)(t)\\
&=&D_n(x_{j0}).
\end{eqnarray*}
Each differential operator of order larger than $1$ appearing in the expression (\ref{D_n}) for $D_n$ acts as zero on $x_{j0}$, so
\begin{eqnarray*}
q_j(n)(0)&=&D_n(x_{j0})\\
&=&\sum_{i\in I}\left(x_{in}\frac{\partial}{\partial x_{i0}}+z_{in}\frac{\partial}{\partial z_{i0}}\right)(x_{j0})\\
&=&x_{jn}.
\end{eqnarray*}
By exactly the same arguments, $p_j(n)(0)=z_{jn}$.\qed

\begin{example}{\em
For $\fg=\mathfrak{sl}_2(\mathbb{R})$, we take the normalisation $q_n(t)=\frac{1}{\sqrt{2}}q_1(n)(t)$ and $p_n(t)=\frac{1}{\sqrt{2}}p_1(n)(t)$.  When $N=0$, factorisation no longer requires inversions in the ring of formal series, and Equation (\ref{Lax-solution2}) becomes
$$e^{q_0(t)}=\frac{e^{q_0(0)}}{\cosh{t\beta}-\sinh{t\beta}},$$
so 
$$q_0(t)=x_0-\log(\cosh{t\beta}-\frac{z_0}{\beta}\sinh{t\beta}),$$ 
where $x_n=q_n(0)$ and $z_n=p_n(0)$ for all $n$, and $\beta=\sqrt{z_0^2+e^{2x_0}}.$  Explicit general solutions for all $N$ are then readily available using jet transformations and Theorem \ref{jet-theorem}.  For example,
\begin{align*}
q_1(t)&=D_1(q_0(t))=\left(x_1\frac{\partial}{\partial x_0}+z_1\frac{\partial}{\partial z_0}\right)q_0(t)\\
&=x_1+\frac{(z_1-x_1z_0)e^{2x_0}\sinh{t\beta}}{\beta^3\cosh{t\beta}-\beta^2z_0\sinh{t\beta}}-t\left(\frac{(z_0z_1+x_1e^{2x_0})(\beta\sinh{t\beta}-z_0\cosh{t\beta})}{\beta^2\cosh{t\beta}-\beta z_0\sinh{t\beta}}\right).
\end{align*}
When initial velocity is zero, that is, $z_n=0$ for all $n$, this reduces to
\begin{eqnarray*}
q_0(t)&=&x_0-\log\cosh(te^{x_0})\\
q_1(t)&=&x_1-tx_1e^{x_0}\tanh(te^{x_0}).
\end{eqnarray*}
In this setting, the jet transformation
$$D_2=\frac12\left(x_1\frac{\partial}{\partial x_0}+z_1\frac{\partial}{\partial z_0}\right)^2+\left(x_2\frac{\partial}{\partial x_0}+z_2\frac{\partial}{\partial z_0}\right)$$
simplifies to $\displaystyle{\frac12 x_1^2\frac{\partial^2}{\partial x_0^2}+x_2\frac{\partial}{\partial x_0}}$, so
\begin{align*}
q_2(t)&=D_2(q_0(t))\\
&=x_2-\frac12 te^{x_0}\left(2x_2-x_1^2\right)\tanh(t e^{x_0})-\frac12 t^2x_1^2e^{2x_0}\sech^2(te^{x_0}),
\end{align*}
and we recover the formulas in (\ref{N=2-solution}).
}
\end{example}
\begin{example}\label{infinite-Toda-example}{\em
Recall the equations of motion (\ref{Toda-Ainfty}) associated with the infinite lattice:
\begin{equation}\label{infinite-lattice}
\bigddot{q}_i=-e^{q_i-q_{i+1}}+e^{q_{i-1}-q_i}
\end{equation}
for all $i\in\mathbb{Z}$.  Finite and affine versions of these equations in type $A$ can be obtained from (\ref{infinite-lattice}), up to a change in basis, by imposing appropriate boundary conditions.  For these equations, Toda found soliton solutions,
\begin{equation}\label{Toda-solution}
e^{-r_j}=1+\gamma_0^2\sech^2(\kappa_0 j\pm \gamma_0t),
\end{equation}
where $r_j=q_{j+1}-q_j$, $\gamma_0=\sinh\kappa_0$, and $\kappa_0\in\mathbb{R}$, which plays the role of the spring constant in an exponential potential.  He then used a continuum limit to relate (\ref{Toda-solution}) to known solutions of the Korteweg-de Vries equation \cite{toda-wadati73}.  In terms of initial conditions, $\kappa_0=\frac12\cosh^{-1}\left(2e^{x_{00}-x_{10}}-1\right)$, where $q_j(0)=x_{j0}$.  

The Takiff version of these equations of motion (\ref{Takiff-Ainfty})
can be solved by applying jet transformations
$$D_n=\left(\exp\sum_{k=1}^N\left(\sum_{i\in\mathbb{Z}} x_{ik}\frac{\partial}{\partial x_{i0}}+z_{ik}\frac{\partial}{\partial z_{i0}}\right)v^k\right)_n$$
to (\ref{Toda-solution}).  The action of $D_n$ on $r_j=r_j(0)(t)$ is well defined, since the only $x_{i0}$ and $z_{i0}$ that appear in the right-hand side of (\ref{Toda-solution}) are $x_{00}$ and $x_{10}$.  That is,
$$r_j(n)=q_{j+1}(n)-q_j(n)=D_n(q_{j+1}(0)-q_j(0))=D_n(r_j(0)).$$
For example, $D_1(e^{-r_j(0)})=-r_j(1)e^{-r_j(0)},$ so we have solutions
\begin{eqnarray*}
e^{-r_j(0)}&=&1+\gamma_0^2\sech^2(\kappa_0j\pm\gamma_0t)\\
-r_j(1)e^{-r_j(0)}&=&D_1(1+\gamma_0^2\sech^2(\kappa_0j\pm\gamma_0t))\\
&=&2\gamma_0\gamma_1\sech^2(\kappa_0 j\pm \gamma_0t)+2\gamma_0^2(\kappa_1 j\pm\gamma_1 t)\sech^2(\kappa_0j\pm \gamma_0 t)\tanh(\kappa_0 j\pm\gamma_0t),
\end{eqnarray*}
where 
\begin{eqnarray*}
\kappa_1&=&D_1\kappa_0=\frac{x_{01}-x_{11}}{2\sqrt{1-e^{x_{10}-x_{00}}}}\\
\gamma_1&=&D_1\gamma_0=\kappa_1\cosh{\kappa_0}=\frac{(x_{01}-x_{11})e^{x_{00}-x_{10}}}{2\sqrt{e^{x_{00}-x_{10}}-1}}.
\end{eqnarray*}
}
\end{example}

  \section{Vinberg quantization}

  For any Lie algebra $\fm$, let
  $$0=U_{-1}(\fm)\subset U_0(\fm)\subset U_1(\fm)\subset \cdots$$
  be the PBW filtration of its enveloping algebra $U(\fm)$.  The associated graded algebra
  $$\hbox{gr}\,U(\fm)=\bigoplus_{n=0}^\infty U_n(\fm)/U_{n-1}(\fm)$$
  is then isomorphic to the symmetric algebra $S(\fm)$. For finite dimensional simple Lie algebras $\fg$, Vinberg \cite{vinberg} asked whether there are commutative subalgebras $A_\mu\subset U(\fg)$ which quantize the (Poisson-commutative) Mischenko-Fomenko subalgebras $B_\mu\subset S(\fg)$ in the sense that $\hbox{gr}\,A_\mu=B_\mu$.  This is sometimes called {\em Vinberg's Problem}.

  The analogous problem in our context is to quantize the Poisson-commutative subalgebras generated by the integrable systems introduced in Section 3.  As a corollary, this will give a second proof of commutativity for the classical integrable systems described in Section 3.  Recall that the integrals of motion $f_{k\ell}=\tr_k(L^\ell)$ are restrictions of functions in $S(\fg_N^R)$ to the coadjoint orbit $\mathcal{O}=\mathcal{O}_{x'}$.  As Poisson algebras, $S(\fg_N^R)|_{\mathcal{O}}=S(\mathfrak{c}_N^R),$ where $\mathfrak{c}_N^R$ is the Lie algebra
  \begin{equation}\label{c_N^R}
    \mathfrak{c}_N^R=\hbox{Span}\{h_i(n),e_\ga(n)\ :\ 1\leq i\leq s,\ \ga\in\Delta,\ 0\leq n\leq N\},
    \end{equation}
  with brackets
  \begin{align}
    [h_i(m),h_j(n)]_R&=[e_\ga(m),e_\gb(n)]_R=0\nonumber\\
    [h_i(m),e_\ga(n)]_R&=\frac12\ga(h_i)e_\ga(m+n),\label{c_N^R-relations}
    \end{align}
  for $1\leq i,j\leq s$, $\ga,\gb\in\Delta$, and $0\leq m,n\leq N$, where $e_\ga(m+n)$ is defined to be $0$ for $m+n>N$.  Viewing the conserved quantities $f_{k\ell}$ as elements of the symmetric algebra $S(\mathfrak{c}_N^R)=\bigoplus_{n=0}^\infty U_n(\mathfrak{c}_N^R)/U_{n-1}(\mathfrak{c}_N^R)$, Vinberg's Problem is to construct commuting elements $u_{k\ell}\in U_k(\mathfrak{c}_N^R)$ so that their images
  $$u_{k\ell}+U_{\ell-1}(\mathfrak{c}_N^R)\in U_\ell(\mathfrak{c}_N^R)/U_{\ell-1}(\mathfrak{c}_N^R)\subset S(\mathfrak{c}_N^R)$$
  are precisely the functions $f_{k\ell}$.  As the quantization of $H=\frac12\tr_N(L^2)$, the element $\frac12 u_{N2}$ can then be regarded as a quantum Hamiltonian, with $\{u_{k\ell}\ :\ 0\leq k\leq N,\ \ell-1\in E(\fg)\}$ as its set of quantum first integrals of motion.

 As in Section 1, we fix a faithful representation $\rho:\ \fg\rightarrow\mathfrak{gl}(M)$ of minimal dimension $M$, and recall that $(h_i|h_j)=\delta_{ij},$ $(e_\ga|e_\gb)=(f_\ga|f_\gb)=0$, and $(e_\ga|f_\gb)=\delta_{\ga\gb}$, with respect to trace form $(x|y)=\tr(\rho(x)\rho(y))$.  The following proposition is a reformulation of a recent result of Molev \cite[Corollary 2.3]{molev}. 
 \begin{proposition}
   Let
   $$F(v)=\sum_{n=0}^N\left(\sum_{i=1}^s h_i(N-n)\ot\rho(h_i)+\sum_{\ga\in\Phi_+}e_\ga(N-n)\ot\rho(f_\ga)+f_\ga(N-n)\ot\rho(e_\ga)\right)v^n,$$
   viewed as an $M\times M$ matrix with entries in $U(\fg_N)\ot K[v]$.  Then the coefficient $(\tr\,F(v)^\ell)_k$ of $v^k$ in $\tr(F(v)^\ell)$ is in the centre of $U(\fg_N)$ for all $0\leq k\leq N$ and $\ell-1\in E(\fg)$.\qed
 \end{proposition}

 \bigskip

 Consider the projection $\pi:\ \fg_N\rightarrow\mathfrak{b}_N$ with respect to the vector space decomposition $\fg_N=\mathfrak{k}_N\oplus\mathfrak{b}_N$, where
 \begin{align*}
   \mathfrak{k}_N&=\hbox{Span}\{e_\ga(n)-f_\ga(n)\ :\ \ga\in\Phi_+,\ 0\leq n\leq N\}\\
   \mathfrak{b}_N&=\hbox{Span}\{h_i(n),\ e_\ga(n)\ :\ 1\leq i\leq s,\ \ga\in\Phi_+,\ 0\leq n\leq N\}.
 \end{align*}
 The map $\pi$ induces a projection $\tilde{\pi}:\ U(\fg_N)=\mathfrak{k}_NU(\fg_N)\oplus U(\mathfrak{b}_N)\rightarrow U(\mathfrak{b}_N),$ which is easily seen to be an associative algebra homomorphism when restricted to the centre $Z(\fg_N)$ of $U(\fg_N)$.  Indeed, for any $z,w\in Z(\fg_N)$, write $z=z_\mathfrak{k}+z_{\mathfrak{b}}$ and $w=w_\mathfrak{k}+w_\mathfrak{b},$ where $z_\mathfrak{k},w_\mathfrak{k}\in\mathfrak{k}_NU(\fg_N)$ and $z_\mathfrak{b},w_\mathfrak{b}\in U(\mathfrak{b}_N)$.  Then
 \begin{align*}
   \tilde{\pi}(zw)&=\tilde{\pi}(wz)=\tilde{\pi}(w_\mathfrak{k}z)+\tilde{\pi}(w_\mathfrak{b}z)\\
   &=\tilde{\pi}(zw_\mathfrak{b})=\tilde{\pi}(z_\mathfrak{k}w_\mathfrak{b})+\tilde{\pi}(z_\mathfrak{b}w_{\mathfrak{b}})\\
   &=\tilde{\pi}(z_\mathfrak{b}w_\mathfrak{b})=z_\mathfrak{b}w_\mathfrak{b}\\
   &=\tilde{\pi}(z)\tilde{\pi}(w).
   \end{align*}

 Let $\mathfrak{n}_N=\hbox{Span}\{e_\ga(n)\ :\ \ga\in\Phi_+,\ 0\leq n\leq N\}$.  Then $[\mathfrak{n}_N,\mathfrak{n}_N]$ is an ideal of $\mathfrak{b}_N$, and we let $\mathfrak{c}_N$ be the Lie algebra $\mathfrak{b}_N/[\mathfrak{n}_N,\mathfrak{n}_N]$.  The quotient map $\phi:\ \mathfrak{b}_N\rightarrow\mathfrak{c}_N$ and rescaling map $r:\ \mathfrak{c}_N\longrightarrow \mathfrak{c}_N^R$ are Lie algebra homomorphisms, where $\mathfrak{c}_N^R$ is the Lie algebra defined in (\ref{c_N^R}) and 
 \begin{align*}
  r:\  &h_i(n)+[\mathfrak{n}_N,\mathfrak{n}_N]\longmapsto 2h_i(n)\\
   &e_\ga(n)+[\mathfrak{n}_N,\mathfrak{n}_N]\longmapsto 2e_\ga(n),\ \hbox{for}\ 1\leq i\leq s,\ \ga\in\Delta,\ \hbox{and}\ 0\leq n\leq N.
   \end{align*}
 They induce enveloping algebra homomorphisms $\tilde{\phi}:\ U(\mathfrak{b}_N)\rightarrow U(\mathfrak{c}_N)$ and $\tilde{r}:\ U(\mathfrak{c}_N)\rightarrow U(\mathfrak{c}_N^R),$ so the composition $\tilde{r}\circ\tilde{\phi}\circ\tilde{\pi}:\ Z(\fg_N)\rightarrow U(\mathfrak{c}_N^R)$ is a homomorphism of associative algebras.  In particular, its image is commutative.

 \begin{theorem}\label{quantum-involution-thm}
   Let
   $$G(v)=\sum_{n=0}^N\left(\sum_{i=1}^s h_i(N-n)\ot\rho(h_i)+\sum_{\ga\in\Delta}e_\ga(N-n)\ot\rho(e_\ga+f_\ga)\right)v^n.$$
   Then
   $$\{u_{k\ell}\ :\ 0\leq k\leq N,\ \ell-1\in E(\fg)\}$$
   is a maximal set of commuting algebraically independent elements of $U(\mathfrak{c}_N^R)$, where $u_{k\ell}=(\tr\, G(v)^\ell)_k.$
Moreover,
   $u_{k\ell}+U_{\ell-1}(\mathfrak{c}_N^R)=f_{k\ell},$
   as elements of $U_\ell(\mathfrak{c}_N^R)/U_{\ell-1}(\mathfrak{c}_N^R)$ in $S(\mathfrak{c}_N^R)=S(\fg_N^R)|\mathcal{O}$. 
   \end{theorem}

\noindent
\proof\ \  That $\{u_{k\ell}\ :\ 0\leq k\leq N,\ \ell-1\in E(\fg)\}$ is a set of commuting elements follows from the discussion above since $Z(\fg_N)$ is manifestly commutative, and
$$u_{k\ell}=\tilde{r}\circ\tilde{\phi}\circ\tilde{\pi}\left(\frac{1}{2^\ell}(\tr\,F(v)^\ell)_k\right)$$
is in the image of the algebra homomorphism $\tilde{r}\circ\tilde{\phi}\circ\tilde{\pi}:\ Z(\fg_N)\rightarrow U(\mathfrak{c}_N^R)$.  As functions on $\mathcal{O}$,
\begin{align*}
  h_i(N-n):\ &L\longmapsto (h_i(N-n)|L)_N=y_i(n)\\
  e_\ga(N-n):\ &L\longmapsto (e_\ga(N-n)|L)_N=b_\ga(n),
  \end{align*}
for
$$L=\sum_{n=0}^N\left(\sum_{j=1}^s y_j(n)h_j(n)+\sum_{\beta\in\Delta}b_\gb(n)(e_\gb(n)+f_\gb(n)\right)\in\mathcal{O},$$
so $h_i(N-n)$ and $e_\ga(N-n)$ coincide with the coordinate functions $y_i(n)$ and $b_\ga(n)$, respectively, as functions on $\mathcal{O}$.  Therefore,
\begin{align*}
  u_{k\ell}+U_{\ell-1}(\mathfrak{c}_N^R)&=\left(\tr\left(\sum_{n=0}^N(\sum_{i=1}^s y_i(n)v^n\ot\rho(h_i)+\sum_{\ga\in\Delta}b_\ga(n)v^n\ot\rho(e_\ga+f_\ga))\right)^\ell\right)_k\\
  &=\tr_k\left(\sum_{n=0}^N\left(\sum_{i=1}^s y_i(n)h_i(n)+\sum_{\ga\in\Delta}b_\ga(n)(e_\ga(n)+f_\ga(n)\right)\right)^\ell\\
  &=f_{k\ell}.
  \end{align*}
Algebraic independence then follows from independence of the polynomials $f_{k\ell}$ in Theorem \ref{system-is-integrable-thm}(ii), for $0\leq k\leq N$ and $\ell-1\in E(\fg)$.  The maximal number of such independent elements is half the dimension of the symplectic manifold $\mathcal{O}$, that is, $\frac12(2(N+1)s)=s(N+1)$, by Proposition \ref{coadjoint-orbit-description-proposition}.  By Chevalley's Restriction Theorem \cite[chapitre VIII, \S8.3, th\'eor\`eme 1] {bourbaki-Lie7-8}, the number of exponents of $\fg$ is its rank $s$, so $\{u_{k\ell}\ :\ 0\leq k\leq N,\ \ell-1\in E(\fg)\}$ is maximal.\qed

\bigskip

As a corollary, we obtain a second proof of commutativity for the classical integrable systems $\mathbb{I}_N(\fg)$ of Theorem \ref{system-is-integrable-thm}.

\begin{corollary}
The functions $f_{k\ell}=\tr_k(L^\ell)$ are mutually commutative, for $k=0,\ldots,N$ and $\ell-1\in E(\fg)$.
  \end{corollary}

\noindent
\proof\ \ This follows from commutativity of the elements $u_{k\ell}\in U(\mathfrak{c}_N^R)$ and the fact that $f_{k\ell}=u_{k\ell}+U_\ell(\mathfrak{c}_N^R)$ as elements of the graded component $S^\ell(\mathfrak{c}_N^R)=U_\ell(\mathfrak{c}_N^R)/U_{\ell-1}(\mathfrak{c}_N^R)$ of $S(\mathfrak{c}_N^R)=\bigoplus_{j=0}^\infty S^j(\mathfrak{c}_N^R)$, together with the following well known result for any Lie algebra $\mathfrak{m}$ (see \cite[Remarques 2.8.7]{dixmier}, for example):

For all $f\in S^m(\mathfrak{m})$, and $g\in S^n(\mathfrak{m})$, let $\tilde{f}\in U_m(\mathfrak{m})$ and $\tilde{g}\in U_n(\mathfrak{m})$ such that
\begin{align*}
  \tilde{f}+U_{m-1}(\mathfrak{m})&=f\\
  \tilde{g}+U_{n-1}(\mathfrak{m})&=g,
\end{align*}
as elements of $S^m(\mathfrak{m})=U_m(\mathfrak{m})/U_{m-1}(\mathfrak{m})$ and $S^n(\mathfrak{m})=U_n(\mathfrak{m})/U_{n-1}(\mathfrak{m})$, respectively.  Then $[\tilde{f},\tilde{g}]=\tilde{f}\tilde{g}-\tilde{g}\tilde{f}\in U_{m+n-1}(\mathfrak{m})$, and $\{f,g\}=[\tilde{f},\tilde{g}]+U_{m+n-2}(\mathfrak{m})$, as elements of $S^{m+n-1}(\mathfrak{m}).$\qed

\appendix

\section{Appendix}

The following easy lemma was used in Section 4.2.  As usual, we write $a(u)_i$ or simply $a_i$ for the coefficient of $u^i$ in the expression  $a(u)=a_0+a_1u+\cdots+a_N u^N +\langle u^{N+1}\rangle$ of any element $a(u)$ in the algebra $K[u]/\langle u^{N+1}\rangle$ of truncated currents over an arbitrary field $K$ of characteristic zero.

\begin{lemma}\label{invertability-and-sqrt}  Let $a(u)$ be a nonzero element of $K[u]/\langle u^{N+1}\rangle$.
\begin{enumerate}
\item[{\rm (i)}] Then $a(u)$ is invertible if and only if $a_0\neq 0$.
\item[{\rm (ii)}] The element $a(u)$ has a square root $b(u)$ in $K[u]/\langle u^{N+1}\rangle$ if and only if $a_k$ has a square root in $K$, where $k=\hbox{min}\{i\geq 0\,:\,a_i\neq 0\}$.
\end{enumerate}
\end{lemma}
\proof\ \ (i) If $a_0=0$, then clearly $(a(u)b(u))_0=0$ for all $b(u)\in K[u]/\langle u^{N+1}\rangle$, and $a(u)$ is not invertible.  If $a_0\neq 0$, then we induct downwards on the minimal positive degree $m=\min\{i>0\,:\,a_i\neq 0\}$ of $a(u)$.  By convention, we define $m=N+1$ if $a(u)=a_0$.  If $m>N$, the result is clear.  Otherwise, $a(u)\big(1-\frac{a_m}{a_0}u^m\big)$ has larger minimal positive degree, and the result holds by induction.

\medskip

\noindent
(ii) If $b(u)^2=a(u)$, then clearly $a_k=b_\ell^2$, where $k=\min\{i\geq 0\,:\,a_i\neq0\}$ and $\ell=\min\{i\geq 0\,:\,b_i\neq 0\}$.

Conversely, if $a_k$ has a square root $s\in K$, we induct on $N$.  When $N=0$, the result is obvious.  By the induction hypothesis, for any $N-1\geq 0$, there exists $r(u)\in K[u]$ such that $r(u)^2+\langle u^N\rangle=a(u)+\langle u^N\rangle.$  Let $n=\min\{i\geq 0\,:\,r_i\neq 0\}$.  If $n=N/2$, then $r(u)^2+\langle u^{N+1}\rangle=r_n^2u^N+\langle u^{N+1}\rangle$, so $a_i=0$ for all $i<N$ and we can take $b(u)=su^n+\langle u^{N+1}\rangle.$  Otherwise, let $c=\frac{a_N-\big(r(u)^2\big)_N}{2r_n},$
and set $b(u)=r(u)+cu^{N-n}+\langle u^{N+1} \rangle.$  
Then $b(u)^2=a(u)$ in $K[u]/\langle u^{N+1}\rangle$.
\qed

\bigskip
\bigskip

\noindent
{\bf Acknowledgements.}  The author is grateful to Carlos Tomei, Rukmini Dey, and Xiao He for many enjoyable discussions.  He also thanks IMPA (Rio de Janeiro) and the University of Alberta, where parts of this project were completed.




\begin{thebibliography}{}

\bibitem[Ara]{arakawa} T.~Arakawa, Representation theory of $W$-algebras and Higgs branch conjecture, in: Proceedings of the International Congress of Mathematicians--Rio de Janeiro 2018.  vol.~II. World Sci.~Publ., Hackensack, NJ, 2018, pp.~1263--1281.


\bibitem[BBT]{Babelon-Bernard-Talon03}
O.~Babelon, D.~Bernard, and M.~Talon, {Introduction to Classical Integrable Systems}, Cambridge, 2003.  

\bibitem[BR]{BR} A.~Babichenko and D.~Ridout, Takiff superalgebras and conformal field theory, J.~Phys.~A {\bf 46} (2013), 125204, 26 pp.
  

\bibitem[Bog]{bogoyavlensky76} O.I.~Bogoyavlensky, On perturbations of the periodic Toda lattice, Commun.~Math.~Phys. {\bf 51} (1976), 201--209.




\bibitem[Bou]{bourbaki-Lie7-8} N.~Bourbaki, \'El\'ements de math\'ematique: Groupes et alg\`ebres de Lie, Chapitres 7 et 8, Hermann, Paris, 1975.





\bibitem[CO]{Casati-Ortenzi06}
  P.~Casati and G.~Ortenzi, New integrable hierarchies from vertex operator representations of polynomial Lie algebras, J.~Geom.~Phys. {\bf 56} (2006), 418--449.
  


\bibitem[CP]{Chari-Pressley94}
V.~Chari and A.~Pressley, {A Guide to Quantum Groups}, Cambridge, 1994.  







\bibitem[Dix]{dixmier}
  J.~Dixmier, {Alg\`ebres enveloppantes}, \'Editions Gauthier-Villars, Paris, 1974. 

\bibitem[FPU]{FPU} E.~Fermi, J.R.~Pasta, S.M.~Ulam, Studies of nonlinear problems, in: Enrico Fermi Collected Papers, Vol. II, Univ.~Chicago Press, Chicago, 1965, pp.~977--988.

\bibitem[Fla]{flaschka} H.~Flaschka, The Toda lattice. I. Existence of integrals, Phys.~Rev.~B {\bf 9} (1974), 1924--1925.

\bibitem[Hen]{henon}M.~H\'enon, Integrals of the Toda lattice, Phys.~Rev.~B {\bf 9} (1974), 1921--1923.

\bibitem[Kos]{kostant79} B.~Kostant, The solution to a generalized Toda lattice and representation theory, Adv.~Math. {\bf 34} (1979), 195--338.




\bibitem[LT]{LT} L.-C.~Li and C.~Tomei, The complete integrability of a Lie-Poisson system proposed by Bloch and Iserles, Int. Math.
Res. Not. (2006), Art. ID 64949, 19 pp.





\bibitem[Mol]{molev} A.I.~Molev, Casimir elements and Sugawara operators for Takiff algebras, J.~Math.~Phys. {\bf 62} (2021), no.~1, paper no.~011701. 

\bibitem[MY]{MY} A.~Moreau and R.~Yu, Jet schemes of the closure of nilpotent orbits, Pacific J.~Math. {\bf 281} (2016), 137--183.


\bibitem[Tod]{toda67} M.~Toda, Wave propagation in anharmonic lattices, J.~Phys.~Soc.~Japan {\bf 23} (1967), 501--506.

\bibitem[TW]{toda-wadati73} M.~Toda and M.~Wadati, A soliton and two solitons in an exponential lattice and related equations, J.~Phys.~Soc.~Japan {\bf 34} (1973), 18--25.

\bibitem[Vin]{vinberg} E.B.~Vinberg, On certain commutative subalgebras of a universal enveloping algebra (Russian), Izv.~Akad.~Nauk SSSR Ser.~Mat. {\bf 54} (1990), 3--25; translation in Math.~USSR-Izv. {\bf 36} (1991), 1--22.

  
\end{thebibliography}
\end{document}